\documentstyle[prd,aps]{revtex}

\input{epsf}

\tightenlines

\newcommand{\fbs}{f_{B{\scriptscriptstyle{s}}}} 
\newcommand{\fb}{f_{B{\scriptscriptstyle{d}}}} 
\newcommand{\fds}{f_{D{\scriptscriptstyle{s}}}} 
\newcommand{\fd}{f_{D{\scriptscriptstyle{d}}}}

\newcommand{\nf}{N_f}

\newcommand{\MeV}{\mbox{MeV}}
\newcommand{\GeV}{\mbox{GeV}}
\newcommand{\fBs}{f_{B\scriptscriptstyle{s}}}
\newcommand{\fBd}{f_{B\scriptscriptstyle{d}}}
\newcommand{\fDs}{f_{D\scriptscriptstyle{s}}}
\newcommand{\fDd}{f_{D\scriptscriptstyle{d}}}

\def\lsim{\raise0.3ex\hbox{$<$\kern-0.75em\raise-1.1ex\hbox{$\sim$}}}
\def\gsim{\raise0.3ex\hbox{$>$\kern-0.75em\raise-1.1ex\hbox{$\sim$}}}

\begin{document}


\preprint{UTCCP-P-68} 

\draft

\title{Decay constants of $B$ and $D$ mesons
from improved relativistic lattice QCD with 
two flavours of sea quarks}

\author{CP-PACS Collaboration\\
  $^1$A.~Ali Khan, $^2$S.~Aoki, $^{1,2}$R.~Burkhalter,
  $^1$S.~Ejiri, $^3$M.~Fukugita, $^4$S.~Hashimoto,
  $^{1,2}$N.~Ishizuka, $^{1,2}$Y.~Iwasaki,
  $^{1,2}$K.~Kanaya, $^4$T.~Kaneko, $^4$Y.~Kuramashi,
  $^1$T.~Manke%
  \thanks{Present address: Department of Physics, Columbia University, 
	538 West 120$^{th}$ St., New York, NY 10027, U.S.A.}, 
  $^1$K.~Nagai, $^4$M.~Okawa,
  $^{1}$H.~P.~Shanahan%
  \thanks{Present address: Department  of Biochemistry and Molecular Biology, 
   	University College London, Darwin Building, Gower St., London, 
	WC1E 6BT, England, U.K.},  
  $^{1,2}$A.~Ukawa,
  $^{1,2}$T.~Yoshi\'e
  } 

\address{
  $^1$Center for Computational Physics,
  University of Tsukuba, Tsukuba, Ibaraki 305-8577, Japan \\
  $^2$Institute of Physics, University of
  Tsukuba, Tsukuba, Ibaraki 305-8571, Japan \\
  $^3$Institute for Cosmic Ray Research,
  University of Tokyo, Tanashi, Tokyo 188-8502, Japan \\
  $^4$High Energy Accelerator Research Organization (KEK),
  Tsukuba, Ibaraki 305-0801, Japan 
}

\date{\today}

\maketitle

\begin{abstract}
  We present a calculation of the $B$ and $D$ meson decay constants 
  in lattice QCD with two ($\nf=2$) flavours of light dynamical quarks, 
  using an $O(a)$-improved Wilson action for both light 
  and heavy quarks and a renormalization-group improved gauge action.
  Simulations are made at three values of lattice spacing 
  $a=0.22, 0.16, 0.11$~fm and four values of sea quark mass in the 
  range $m_{PS}/m_V\approx 0.8$--$0.6$.
  Our estimate for the continuum values of the decay constants are 
  $\fBd = 208(10)(11)\MeV,
   \fBs = 250(10)(13)(^{+8}_{-0})\MeV,
   \fDd = 225(14)(14)\MeV,
   \fDs = 267(13)(17)(^{+10}_{-0})\MeV$ for $N_f=2$ where 
  the statistical and systematic errors are separately listed, and the 
  third error for $\fBs$ and $\fDs$ show uncertainty of determination of 
  strange quark mass. We also carry out a set of quenched simulations 
  using the same action to make a direct examination of sea quark effects. 
  Taking the ratio of results for $\nf=2$ and $N_f=0$, we obtain
 $\fb^{\nf=2}/\fb^{\nf=0} = 1.11(6), 
  \fbs^{\nf=2}/\fbs^{\nf=0} = 1.14(5),
  \fd^{\nf=2}/\fd^{\nf=0} = 1.03(6),
  \fds^{\nf=2}/\fds^{\nf=0} = 1.07(5)$. 
They show a 10--15\% increase in the $\nf=2$ results over those of $N_f=0$
for the $B$ meson decay constants,  
while evidence for such a trend is statistically less clear 
for the $D$ meson decay constants.
\end{abstract}
\pacs{PACS: 12.15.Hh, 12.38.Gc, 13.20.-v}

\section{Introduction}
The accurate determination of the CKM matrix elements is
one of the most important tasks of flavour physics.
The Standard Model prediction of the unitarity of the matrix
still has to be tested, especially for the unitarity
relation involving the most off-diagonal elements, which
contain the source of the CP violation in the Standard
Model.

Two of the matrix elements,  $|V_{td}|$ and $|V_{ts}|$, can be
extracted from experimental data of the oscillation
frequency $\Delta m_q$ of $B_q-\overline{B}_q$ systems ($q$
denotes either $d$ or $s$ quark) through the relation
\cite{Buras:1990fn} 
\begin{equation}
  \label{eqn:bbbar-osc}
  \Delta m_q =
  \frac{G_F^2}{6\pi^2} M_W^2 S_0(x_t) 
  \eta_B M_{B_q}
  f_{B_q}^2 \hat{B}_{B_q} |V_{tq} V^*_{tb}|^2,
\end{equation}
where the factors other than $f_{B_q}^2 \hat{B}_{B_q}$ are
known either experimentally or through perturbative calculations 
in QCD. The non-perturbative coefficients $f_{B_q}$ and
$\hat{B}_{B_q}$ are defined as
\begin{equation}
  \label{eq:f_B_definition}
  \langle 0 | \bar{b}\gamma_{\mu}\gamma_5 q |B_q(p) \rangle
  = i f_{B_q} p_{\mu},
\end{equation}
and
\begin{equation}
  \label{eq:B_B_definition}
  \hat{B}_{B_q} = R_B(\mu)
  \frac{\langle \bar{B}_q | 
    \bar{b} \gamma_{\mu}(1-\gamma_5) q
    \bar{b} \gamma_{\mu}(1-\gamma_5) q
    |B_q \rangle}{
    \frac{8}{3} f_{B_q}^2 M_{B_q}^2
    },
\end{equation}
where  $R_B(\mu)$ denotes a renormalization group factor to
eliminate the variation due to the scale $\mu$ where the
four-quark operator 
$\bar{b} \gamma_{\mu}(1-\gamma_5) q
 \bar{b} \gamma_{\mu}(1-\gamma_5) q$
is defined.
In this paper we shall focus on the decay constants
$f_{B_q}$, leaving the bag parameter $\hat{B}_{B_q}$ for
future studies.

Experimentally, the Cabibbo-allowed leptonic decay
$D_s\rightarrow\tau\bar{\nu}_{\tau}$ has been measured and 
the recent values for $f_{D_s}$ are
285$\pm$20$\pm$40 MeV (ALEPH \cite{ALEPH}) and
280$\pm$19$\pm$44 MeV (CLEO \cite{CLEO}).
On the other hand, a measurement of the decay constant $f_B$
is difficult, since $B^+\rightarrow l^+\bar{\nu}_l$ 
is Cabibbo-suppressed in the 
Standard Model. Hence $f_{B_q}$ has to be provided from theory, 
while $f_{D_q}$ can be used to check the calculational method. 

The calculation of these decay constants have been
carried out extensively in the quenched approximation in
lattice QCD, where vacuum polarization effects are neglected 
in order to reduce the computational requirements.
A recent summary of these attempts is given in 
Refs.~\cite{Hashimoto:1999bk,Draper_lat98}.
Although the approximation provides a useful first step in a lattice 
QCD determination of the decay constants, 
the size of the resulting systematic error is not clear.
In the quark
potential model,
the decay constant is proportional to
the wave function at the origin squared 
and the potential at such short distance scales can be
expressed in terms of the running coupling constant.
Therefore, one can demonstrate,  on a heuristic level, 
that the decay constant is affected by
the number of flavours $\nf$, where $\nf=0$ corresponds to
the quenched approximation.
Additionally, a recent study of the light hadron
spectrum in the quenched approximation indicates a deviation
of about 10\% from experiment \cite{Aoki:2000yr}.
For the decay constants, quenched chiral perturbation theory 
suggests \cite{Booth:1995hx,Sharpe:1996qp} 
that the deviation introduced by the approximation may be significant.

The elimination of this approximation is numerically highly
intensive and has become realistic only recently.
The MILC collaboration \cite{Bernard:1998xi} and 
Collins \textit{et al.} \cite{Collins:1999ff} have performed
first calculations of the decay constants on the lattice
with two degenerate sea quark flavours, and found an
indication that $f_{B_q}$ is considerably larger in the presence
of sea quarks.
In these studies, however, the discretization of the sea
quarks is defined using the staggered fermion action, which
is different from that used for the light valence quark
(Wilson fermion in Ref.~\cite{Bernard:1998xi} and the
$O(a)$-improved (clover) fermion in Ref.~\cite{Collins:1999ff}).
It would complicate the discretization error in the results.
In fact, the observed $a$, (the lattice spacing), dependence in
$f_B$ found in Ref.~\cite{Bernard:1998xi} is rather different
between quenched and unquenched calculations, even though the 
formulations for valence heavy and light quarks are the
same. 

In our work we apply a consistent formulation where for both 
sea and light valence quarks we use the same action, and study the
$a$ dependence by performing three sets of two-flavour 
calculations at $a$ $\simeq$ 0.22, 0.16 and 0.11 fm.
For comparison, we carry out quenched calculations at
ten different values of $a$ covering the range studied in
the two flavour calculations.
We employ the $O(a)$-improved quark action \cite{ref:clover} 
for both sea and valence light quarks.  The same
action is used for the heavy quark, applying the non-relativistic
reinterpretation of Ref.~\cite{El-Khadra:1997mp}.
The gauge field is described by a renormalization group
improved action \cite{ref:Iwasaki83}, which reduces the
descretization error on the coarse lattices on which our 
calculations are made.
 
The rest of this paper is outlined as follows. 
In Sec.~\ref{sec:action} we discuss the lattice actions 
and the formulation to treat heavy quarks.
The computational details involved in the calculation are
described in Sec.~\ref{sec:detail}, and our analysis procedures 
in Sec.~\ref{sec:analysis}.
We present the results in Sec.~\ref{sec:results} where we discuss 
in particular how we estimate the values in the continuum limit and 
their errors, and make a comparison 
between the $\nf=0$ and $\nf=2$ results. 
Our conclusions are summarized in Sec.~\ref{sec:conclusion}.

\section{Lattice Actions}
\label{sec:action}

\subsection{Light sector}
\label{sec:Lattice_Actions:Light_sector}

The renormalization group (RG) improved gauge action we employ takes the
form \cite{ref:Iwasaki83}
\begin{equation}
  \label{eqn:gauge-action}
  S^{\mathbf{R}}_g = 
  \frac{\beta}{6}
  \left( 
    c_0 \sum W_{1\times 1} + 
    c_1 \sum W_{1\times 2}
  \right),
\end{equation}
where $W_{1\times 1}$ and $W_{1\times 2}$ are the Wilson
loops of size $1\times 1$ and $1\times 2$ respectively, and
the sums run over all possible sites and orientations.
The parameter $\beta$ is related to the bare gauge coupling
$g_0^2$ through $\beta = 6/g_0^2$.
The coefficients $c_0$ and $c_1$ are defined as 
\begin{eqnarray}
  c_0 & = & 3.648,
  \label{eqn:c0} \\
  c_1 & = & \frac{1}{8} \left( 1 - c_0 \right) = -0.331,
  \label{eqn:c1}
\end{eqnarray}
which are chosen so as to approximate the renormalization
group trajectory in two dimensional operator space.

For quarks we employ the $O(a)$-improved (clover) action 
\cite{ref:clover} defined by
\begin{equation}
  S^{\mathbf{C}}_q = 
  \sum_{x,y} \bar{\psi}_x \left[
    D^{\mathbf{W}}_{xy} 
    - c_{SW} K \sum_{\mu < \nu}
    \sigma_{\mu\nu} F_{\mu\nu}
  \right] \psi_y, 
\end{equation}
where $D^{\mathbf{W}}_{xy}$ is the standard Wilson
formulation of the Dirac fermion matrix
\begin{equation}
  D^{\mathbf{W}}_{xy} = \delta_{xy} - K
  \sum_{\mu} \left\{ 
    (1-\gamma_{\mu}) U_{x,\mu} \delta_{x+\hat{\mu},y} 
    + (1+\gamma_{\mu}) U^{\dagger}_{x,\mu}
    \delta_{x,y+\hat{\mu}}
  \right\}
\end{equation}
and the matrix $F_{\mu\nu}$ is the simplest definition of
the field strength, 
\begin{equation}
  F_{\mu\nu} = \frac{1}{8i}
  \left( f_{\mu\nu} - f^\dagger_{\mu\nu}\right),
\end{equation}
where $f_{\mu\nu}$ is the standard clover-shaped definition
of the gauge field strength.
The leading discretization error in the Wilson fermion
action ($c_{SW}=0$) is removed by appropriately tuning the
parameter $c_{SW}$.
We apply a mean field approximation $c_{SW}$ = $P^{-3/4}$,
where $P$=$\langle W_{1\times 1} \rangle$, and perturbative
expansion at one-loop $P = 1 - 0.1402 g_0^2$ is used to
evaluate $P$.
With this choice, the leading contributions among remaining
discretization errors are $O(\alpha_s a)$ and $O(a^2)$ for
light quarks. There is reasonable agreement between $P$ 
measured on the lattice and the above perturbative definition 
with the difference being at worst 8\% 
\cite{CP-PACS_lighthadron}. Furthermore, there is also 
good agreement between the above definition of $c_{SW}$ and
the one-loop computed value \cite{csw-one-loop}.

The efficacy of this choice of actions over the standard
action has been demonstrated in Ref.~\cite{Kaneko} by examining
the rotational invariance of the static potential and the
scaling behaviour of the light hadron spectrum. 
In using the clover fermion action, we also note that
care must be taken in defining currents, which will be
discussed below.

\subsection{Heavy Quarks}

It seems implausible to examine hadrons
containing heavy quarks with mass $m_Q a> 1$ on a lattice with the 
spacing $a$, as one expects the discretization effects to become 
uncontrollably large for such large masses.
However, this is not necessarily true for heavy-light mesons.
The spatial momentum of the light degrees of freedom in the
heavy-light system is controlled by the QCD scale
$\Lambda_{QCD}$ rather than the much larger heavy quark mass scale. 
In the limit of infinite $m_Q$,  the heavy quark mass
decouples from the dynamics of the system, and the Heavy
Quark Effective Theory (HQET) \cite{HQET} becomes a good
approximation.
At a finite $m_Q$, the correction may be incorporated as an
expansion in $1/m_Q$, which is a basis of the
non-relativistic QCD (NRQCD).

On the lattice, it is straightforward to formulate the
static \cite{static} and NRQCD \cite{NRQCD-1,NRQCD-2}
actions, and a number of (quenched) calculations of $f_B$
have been performed using them.
Another formulation to realize the idea of HQET on the
lattice \cite{El-Khadra:1997mp} is also useful, as it uses 
the same relativistic form of the quark action as that for light quarks except
that the \textit{bare} heavy quark mass $m_0$ may be taken
to be arbitrarily large. 

For the heavy-light system, where the typical spatial
momentum of the heavy quark is small compared to the inverse
lattice spacing, one can construct an effective Hamiltonian
starting from a relativistic action, 
\begin{equation}
  \label{eqn:Pauli}
  \hat{H} \approx 
  \hat{\bar{\Psi}} \left[
    M_1 + \gamma_0 A_0 
    - \frac{\mathbf{D}^2}{2 M_2}
    - \frac{i \mathbf{\Sigma}.\mathbf{B}}{2 M_B} 
    -\gamma_0 \frac{[\mathbf{\gamma}\cdot\mathbf{D},
                     \mathbf{\gamma}\cdot\mathbf{E}]
                     }{8 M_E^2}
  \right] \hat{\Psi},
\end{equation}
where $\mathbf{D}$ is the covariant derivative, 
$\mathbf{\Sigma}$ the Pauli spin matrices,
and $\mathbf{B}$ and $\mathbf{E}$ are the 
chromomagentic and chromoelectric fields
respectively
and  an expansion in small spatial momentum or equivalently
in $a\mathbf{D}$ on the lattice is performed.
This Hamiltonian is equivalent to the
standard non-relativistic Hamiltonian if the ``mass''
parameters $M_1$, $M_2$, $M_B$ and $M_E$ are equal to each
other. 
Those are, however, different functions of $am_0$ and not
necessarily equal to each other, unless the parameters in
the initial relativistic action are appropriately tuned.
The strategy suggested in Ref.~\cite{El-Khadra:1997mp} is,
therefore, to take the action as an effective theory to
generate the dynamics described by (\ref{eqn:Pauli}).
The appropriate mass parameter in the non-relativistic
effective theory is the ``kinetic'' mass $M_2$, while the
``pole'' mass $M_1$ does not affect the dynamics of heavy
quark and plays merely a role of energy shift in this
formalism. 
To obtain a correct action at order $1/M$, the mass
parameter which characterizes the spin-magnetic interaction 
$M_B$ must be equal to $M_2$, which is satisfied for the
$O(a)$-improved (clover) action up to perturbative
corrections. 
On the other hand, there is no tunable parameter in the
clover action to make $M_E$ equal to $M_2$ and $M_B$, so
that the contributions of $O(1/M^2)$ and higher are not
correctly described by the clover action.

At tree level, the kinetic mass $M_2$ of the heavy quark
is given by \cite{El-Khadra:1997mp}
\begin{equation}
  \label{eq:M_2}
  aM_2 = \left(   \frac{2}{am_0(2+am_0)} + \frac{1}{1+am_0}
    \right)^{-1},
\end{equation}
where the bare mass $am_0$ is defined as 
$am_0 = \frac{1}{2}(\frac{1}{K}-\frac{1}{K_c})$.
The one-loop relation is also known 
\cite{Mertens_et_al,Kuramashi,AHIO} for
the standard plaquette gauge action but not for our choice of
the action.
Hence we employ the tadpole improvement \cite{LM} of the above
relation, which is obtained by simply replacing $am_0$ with
$8K_c am_0$.

The heavy-light meson mass $aM_{HQET}$ defined in the HQET 
is then obtained as \cite{BLS,Aoki:1998ji}
\begin{equation}
  \label{eq:HQET_mass}
  aM_{HQET}^{M} = aM_{pole}^M + (aM_2^Q - aM_1^Q),
\end{equation}
from the pole mass $aM_{pole}$ extracted from the
exponential fall off of the heavy-light propagator.
The superscript $M$ or $Q$ in (\ref{eq:HQET_mass})
distinguishes the mass of the heavy-light meson ($M$) from
the heavy quark mass ($Q$). The parameters $aM_1^Q$ and 
$aM_2^Q$ are the tree-level defined pole and kinetic masses of the heavy quark. 

An alternative way to obtain the heavy-light meson mass is
to measure its energy-momentum dispersion relation and fit
with the form 
$E({\mathbf{p}}) = M_{pole} + {\mathbf{p}}^2/(2M_{kin}) + 
O( {\mathbf{p}}^4)$
to extract the ``kinetic'' mass $M_{kin}$ (as employed in
\cite{El-Khadra:1998hq}). 
Unfortunately, for the lattices which were used to quote our
final results the statistical ensemble 
was not large enough to obtain an accurate measurement
of $M_{kin}$.
For this reason, this choice of the kinetic
mass will not be further discussed here.

The axial current to be measured should also be modified to
obtain the results correct at $O(1/M)$ according to
\begin{equation}
  \label{eqn:field-redef}
  h \rightarrow 
  ( 1 - a d_1 {\mathbf{\gamma}}\cdot{\mathbf{D}}) h,
\end{equation}
where $h$ is the heavy quark field
and equivalently for $\bar{h}$, and the parameter $d_1$
is a function of $am_0$. 
At the tree level, it is given by \cite{El-Khadra:1997mp} 
\begin{equation}
  \label{eq:d_1}
ad_1 = \frac{1+am_0}{ am_0 (2+am_0)} - \frac{1}{2aM_2}, 
\end{equation}
and the axial vector current for heavy-light mesons, 
correct to $O(1/M)$, takes the form
\begin{equation}
A_\mu(x)= \bar{l}(x) \gamma_5 \gamma_\mu h(x)
	- a d_1\bar{l}(x) \gamma_5 \gamma_\mu
  {\mathbf{\gamma}}\cdot{\mathbf{\Delta}} h(x).
\label{eq:current}
\end{equation}
where $l$ is the  light quark field. 
The tadpole improvement of $d_1$ may be applied again with the
replacement $am_0 \rightarrow 8K_c am_0$.

The following point should also be noted. 
The action being used is still a relativistic action and as
the lattice spacing becomes smaller, it is expected that
theory should smoothly cross over to a fully relativistic
theory.
That means the mass parameters $M$'s become identical as
$am_0$ decreases.
The lattice spacing dependence of physical quantities, such
as $f_B$ and $f_D$, is, however, highly nontrivial unless
$m_0$ is much smaller than $1/a$, and the continuum
extrapolation in such a situation would not be justified
with any simple ansatz, \textit{e.g.} linear or quadratic in
$a$. 
The formulation is, therefore, treated as an effective
theory (like NRQCD), and the discretization error should
be reasonably small 
at fixed $a$ in order to obtain reliable
results. 

Despite the caveat of the preceding paragraph, this
approach has been successfully implemented in the quenched
approximation in Refs.~\cite{Aoki:1998ji,El-Khadra:1998hq} using
the plaquette gauge action. 
Since we use a gauge action which has been unused in
the previous heavy quark calculations it is important for us
that we repeat the calculation in the quenched approximation in
order to see if the quenched results obtained with the ``standard''
plaquette gauge action are reproduced.

\section{Computational Details}
\label{sec:detail}

\subsection{Gauge Fields}
Gauge configurations were generated for $\nf$=0 and $\nf$=2
using the RG improved gauge action and the $O(a)$-improved
Wilson quark action as discussed in
Sec.~\ref{sec:Lattice_Actions:Light_sector}. 
Technical details on the configuration generation for $\nf=2$, carried out 
with the Hybrid Monte Carlo algorithm, are described in 
our dynamical QCD calculations papers
\cite{Burkhalter:1998wu,CP-PACS_lightquark,CP-PACS_lighthadron}.

In the $\nf$=2 calculations, we performed three sets of
calculations at bare gauge couplings $\beta$ = 1.8, 1.95 and
2.1, which correspond to the lattice spacing $a\sim$ 0.22,
0.16 and 0.11 fm respectively.
The lattice size is 12$^3\times$24 ($\beta$=1.8),
16$^3\times$32 (1.95) and 24$^3\times$48 (2.1), with which
the physical volume is approximately (2.5 fm)$^3$.
For each set, we carried out runs at four values of sea
quark mass in order to take the chiral limit of sea quark.
The four sea quark masses are tuned so that the
pseudoscalar-to-vector mass ratio $m_{PS}/m_V$ becomes
roughly 0.80, 0.75, 0.70 and 0.60, which correspond to the
range of quark mass of 3 $\sim$ 0.5 times physical strange
quark mass.
The simulation parameters are listed in Table
\ref{table:lattice-params-nf2}, where the number of HMC
trajectories is also shown.  We note that at $\beta=2.1$ the configurations 
analyzed constitute the first half of the ensemble for each sea 
quark mass.
The full set of configurations is used at $\beta=1.95$ and 1.8. 
The measurements are performed on configurations separated by
10 HMC trajectories at $\beta$ = 1.8 and 1.95 and by 5
trajectories at $\beta$=2.1.
The statistical analysis is done using the jackknife
method in order to take the correlation of successive
trajectories into account.
The bin size is 50 trajectories
for all $\nf=2$ runs, which has been determined to be a suitable length for 
eliminating autocorrelations \cite{CP-PACS_lighthadron}.

The lattice spacing is determined for each $\beta$ value
using the $\rho$ meson mass as input at the physical sea quark
limit.
The chiral extrapolation of light hadrons is discussed in
\cite{Burkhalter:1998wu,CP-PACS_lightquark,CP-PACS_lighthadron}.
The lattice spacings are listed in Table \ref{table:lattice-params-chiral-nf2}.

In order to see the sea quark effect consistently using our
choice of gauge and quark actions, we prepared ten sets of
the quenched ($\nf$=0) gauge configurations.  The values of 
$\beta$ are chosen so that the string 
tension matches with each of full QCD configurations at $\beta=1.95$ 
or 2.1 at four sea quark masses and also in the chiral limit.  
For calculating lattice spacing and hence the physical value of the decay constants,   
the $\rho$ meson mass is used as input in conjunction with the vector masses measured
on the lattice extrapolated to the light quark masses.
The detail of our parameter choice in the quenched runs is
summarized in Table \ref{table:lattice-params-nf0}.

\subsection{Valence quarks}

The heavy and light quark propagators are calculated on each 
set of the gauge configurations for the
$O(a)$-improved Wilson action with the same choice of
$c_{SW}$ as used in the configuration generation.
For each set of gauge configurations, eight values of the
heavy quark mass are chosen so that their HQET mass
(\ref{eq:HQET_mass}) lie roughly on the interval of the $b$
and $c$ quark masses.  

The light quark mass on the dynamical configurations is
the same as their sea quark mass.
In addition, we choose another quark mass for each set
of configurations so that it satisfies $m_{PS}/m_V = 0.688$.
To compute any of the observables at the strange quark mass,
the relevant observables are 
interpolated to the strange quark mass defined from the mass
of the $K$ or $\phi$. 
The light quark masses for $\nf = 0$ are chosen to
take values approximately the same as those for the equivalent lattices
(i.e. those lattices with matched string tension) for $\nf=2$. 

The gauge configurations are fixed to the Coulomb gauge 
with a global maximum residue for $Tr(\partial_i A_i)^2$
set to $10^{-14}$ or less.
The light quark propagators are solved with local sources
while the heavy quark propagators are computed with local
and smeared sources.
The smearing is made with the exponential function
$A\exp(-Br)$, with the mean radius $1/B$ chosen to
approximately reproduce the heavy-light wave function.
The parameters $A$ and $B$ are listed in
Table~\ref{table:smearing-params}. 

Both light and heavy quark propagators are obtained with a solver based 
on the BiCGStab algorithm.  For large values of heavy quark mass, 
stopping the solver if the residue becomes smaller than some minimum is 
not sufficient for obtaining the solution at large time separations. 
In this case, the iteration of the solver is applied a
minimum of $2\times T$ times, where $T$ is the temporal
extent of the lattice, before applying the maximum residue
criterion.

\subsection{Heavy-light current}

We compute the correlation functions constructed from the
following operators
\begin{eqnarray}
  P(x) & = & \bar{l}(x) \gamma_5 h(x), \\
  A(x) & = & \bar{l}(x) \gamma_5 \gamma_0 h(x), \\
  \delta A(x) & = &
  \bar{l}(x) \gamma_5 \gamma_0 
  {\mathbf{\gamma}}\cdot{\mathbf{\Delta}} h(x).
\label{eq:five}
\end{eqnarray}
The heavy and light quark fields $h$ and $l$ are normalized
with $\sqrt{1-3K/4K_c}$, which is motivated with the
nonrelativistic interpretation \cite{El-Khadra:1997mp}
together with the tadpole improvement \cite{LM}.
The derivative current $\delta A$ is used to construct the
modified current according to (\ref{eq:current}), and
$\Delta$ is the discretised covariant derivative
defined as 
\begin{equation}
  \Delta_i h(x) = \frac{1}{2}
  \left[
    U_i(x)                 h(x+\hat{i}) - 
    U^\dagger_i(x-\hat{i}) h(x-\hat{i})
  \right].
\end{equation}

Specifically, we measure the correlation functions 
\begin{equation}
  \begin{array}{cc}
    \sum_{\vec{x}}
    \langle P_L(\vec{x},t) P^\dagger_S(0) \rangle, 
    & 
    \sum_{\vec{x}}
    \langle P_L(\vec{x},t) P^\dagger_L(0) \rangle, 
    \\
    \sum_{\vec{x}}
    \langle A(\vec{x},t) P^\dagger_S(0) \rangle, 
    &
    \sum_{\vec{x}}
    \langle A(\vec{x},t) P^\dagger_L(0) \rangle,
    \\
    \sum_{\vec{x}}
    \langle \delta A(\vec{x},t) P^\dagger_S(0) \rangle,
    &
    \sum_{\vec{x}}
    \langle \delta A(\vec{x},t) P^\dagger_L(0) \rangle,
  \end{array}
\end{equation}
where the subscripts $S$ and $L$ on the pseudoscalar
operators indicate whether smeared or local operators are
employed. 
The axial current $A$ and the derivative current are always
local.

\section{Analysis}
\label{sec:analysis}

\subsection{Correlators}
The correlation functions defined above take the following
form for large Euclidean time separation (we take $a=1$ for
simplicity) 
\begin{eqnarray}
  \label{eq:PP_fit}
  \sum_{\vec{x}} 
  \langle P_L(\vec{x},t) P^\dagger_{(L,S)}(0) \rangle
  & = & 
  \frac{{\cal Z}_{P_L} {\cal Z}_{P_{(L,S)}}}{2M}
  e^{-M_{pole} T/2} \cosh{( M_{pole}(T/2 -t))}
  \nonumber \\
  & + & 
  \frac{{\cal Z}'_{P_L} {\cal Z}'_{P_{(L,S)}}}{2M'}
  e^{-M'_{pole} T/2} \cosh{( M'_{pole}(T/2 - t))},
  \\
  \sum_{\vec{x}}
  \langle A(\vec{x},t) P^\dagger_{(L,S)}(0) \rangle 
  & = & 
  \frac{{\cal Z_A} {\cal Z}_{P_{(L,S)}}}{2M}
  e^{-M_{pole} T/2} \sinh{( M_{pole}(T/2 -t))}
  \nonumber \\
  & + &
  \frac{{\cal Z}'_A {\cal Z}'_{P_{(L,S)}}}{2M'}
  e^{-M'_{pole} T/2} \sinh{( M'_{pole}(T/2 - t))},
  \\
\label{eq:delAP}
  \sum_{\vec{x}}
  \langle \delta A(\vec{x},t) P^\dagger_{(L,S)}(0) \rangle
  & = & 
  \frac{\delta{\cal Z}_A {\cal Z}_{P_{(L,S)}}}{2M}
  e^{-M_{pole} T/2} \sinh{( M_{pole}(T/2 -t))}
  \nonumber \\
  & + &
  \frac{\delta{\cal Z}'_A {\cal Z}'_{P_{(L,S)}}}{2M'}
  e^{-M'_{pole} T/2} \sinh{( M'_{pole}(T/2 - t))},
\end{eqnarray}
where $T$ is the temporal extent of the lattice, and $M$ and
$M'$ are masses of the ground and the first excited
pseudoscalar states respectively. 
The masses extracted from the $t$ dependence of the
correlation functions are the pole masses, while the $M$'s
appearing in the denominator come from the normalization of
states, and their definition need not be specified for calculating the combination
of $f_P \sqrt{M}$.

The matrix elements ${\cal Z}$ are defined as
\begin{eqnarray}
  {\cal Z}_{P_{(L,S)}} & = &
  \langle 0 | P_{(L,S)}(0) | P(\mathbf{0}) \rangle, \\
  {\cal Z}_A & = & \langle 0 | A(0) | P(\mathbf{0}) \rangle, \\
  \delta{\cal Z}_A & = & \langle 0 | \delta A(0) | P(\mathbf{0}) \rangle,
\end{eqnarray}
where $| P(\mathbf{0}) \rangle$ represents the heavy-light
pseudoscalar meson state at rest.
The primed quantities are defined in a similar manner for the
first excited state.

We carry out a simultaneous fit of the three correlators 
(\ref{eq:PP_fit})--(\ref{eq:delAP}). 
Formally, for large Euclidean times, the contribution of the
excited state will be negligible.
However, it is included in the fit so as to use a wider
range of Euclidean times and reduce the size of the
statistical error.
For those sets of configurations with lattice volumes of
size 12$^3\times$ 24 and 16$^3\times$32 the sample size is
large enough to perform a correlated fit of the local and
smeared source data, where the correlation among different
time slices are also taken into account.
For the largest lattice 24$^3\times$48, such a  correlation
matrix appeared to be too large to achieve a stable fit with
our statistics. 
We, therefore, use the uncorrelated fit throughout our
statistical analysis, and check that the results are
unchanged within statistical errors with the correlated fit when it is possible.

The fit criteria we apply for selecting the fit range are as follows 
\cite{Shanahan:1997pk}: 
(i) The quality of fit, $Q$, should be acceptable, {\it e.g.,} $Q>0.1$. 
(ii) The results for the chosen fit range should agree to
  within one standard deviation of the results when the
  minimum time slice is increased or decreased by one
  time-slice. 
(iii) There should be agreement between the ground state
  results obtained using a single-exponential fit and a
  double-exponential fit. 
  This condition increases our confidence that higher state 
  contamination is eliminated.
(iv) In the double-exponential fit the ground and excited
  state energies must be statistically resolvable,
  {\it i.e.}, there must be more than one standard deviation
  between their central values (since we expect the physical states
  to be distinctly seperated). 

\begin{figure}
  \begin{center}
    \leavevmode
    \epsfxsize=0.70 \hsize
    \epsffile{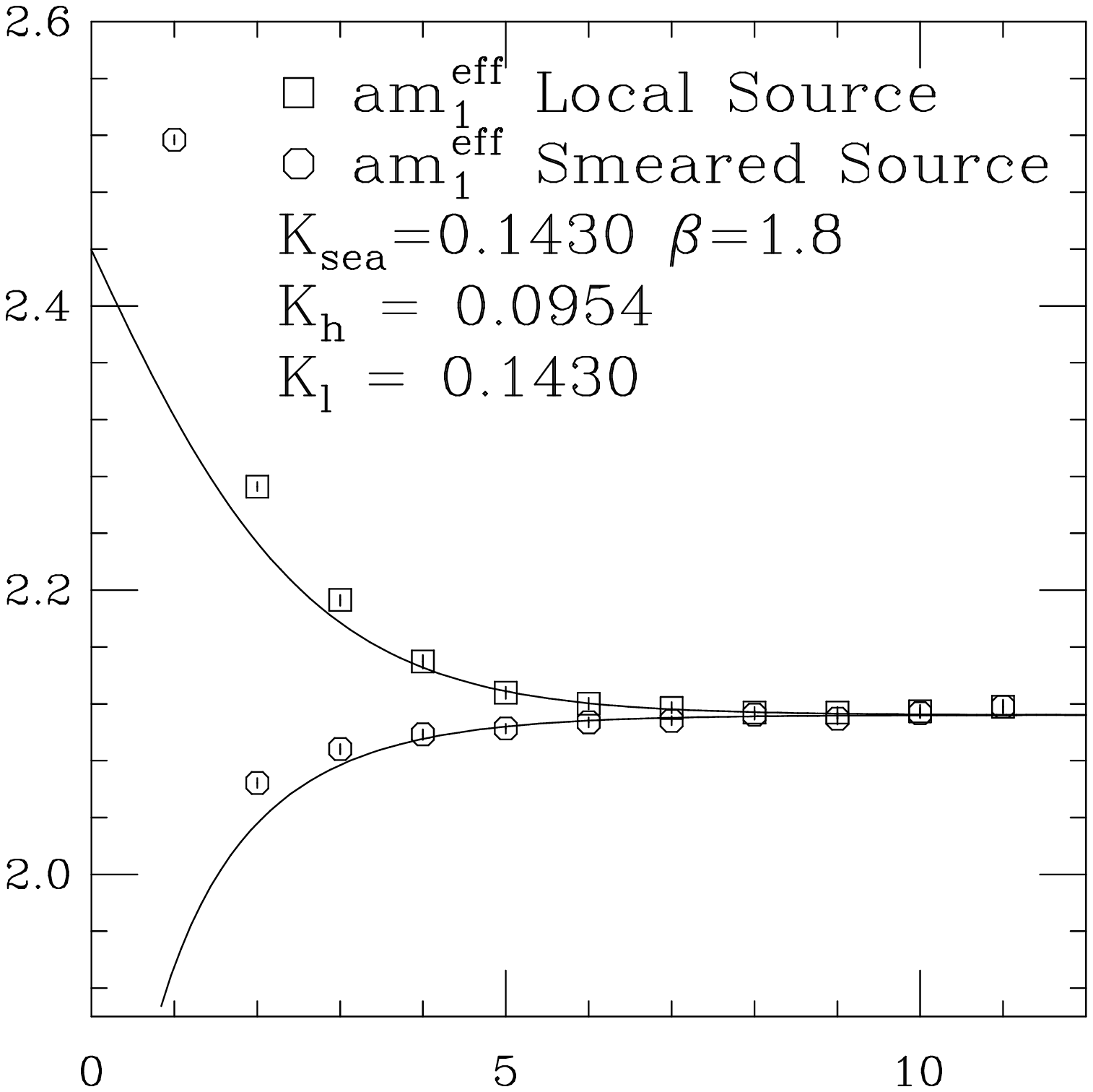}
    \caption{Typical effective mass  plots at $\beta=1.8$ for
      $\nf=2$. 
      The fit range is from 3 to 11.}
  \label{fig:m1eff-B1.8}
  \end{center}
\end{figure}

\begin{figure}
  \begin{center}
    \leavevmode
    \epsfxsize=0.70 \hsize
    \epsffile{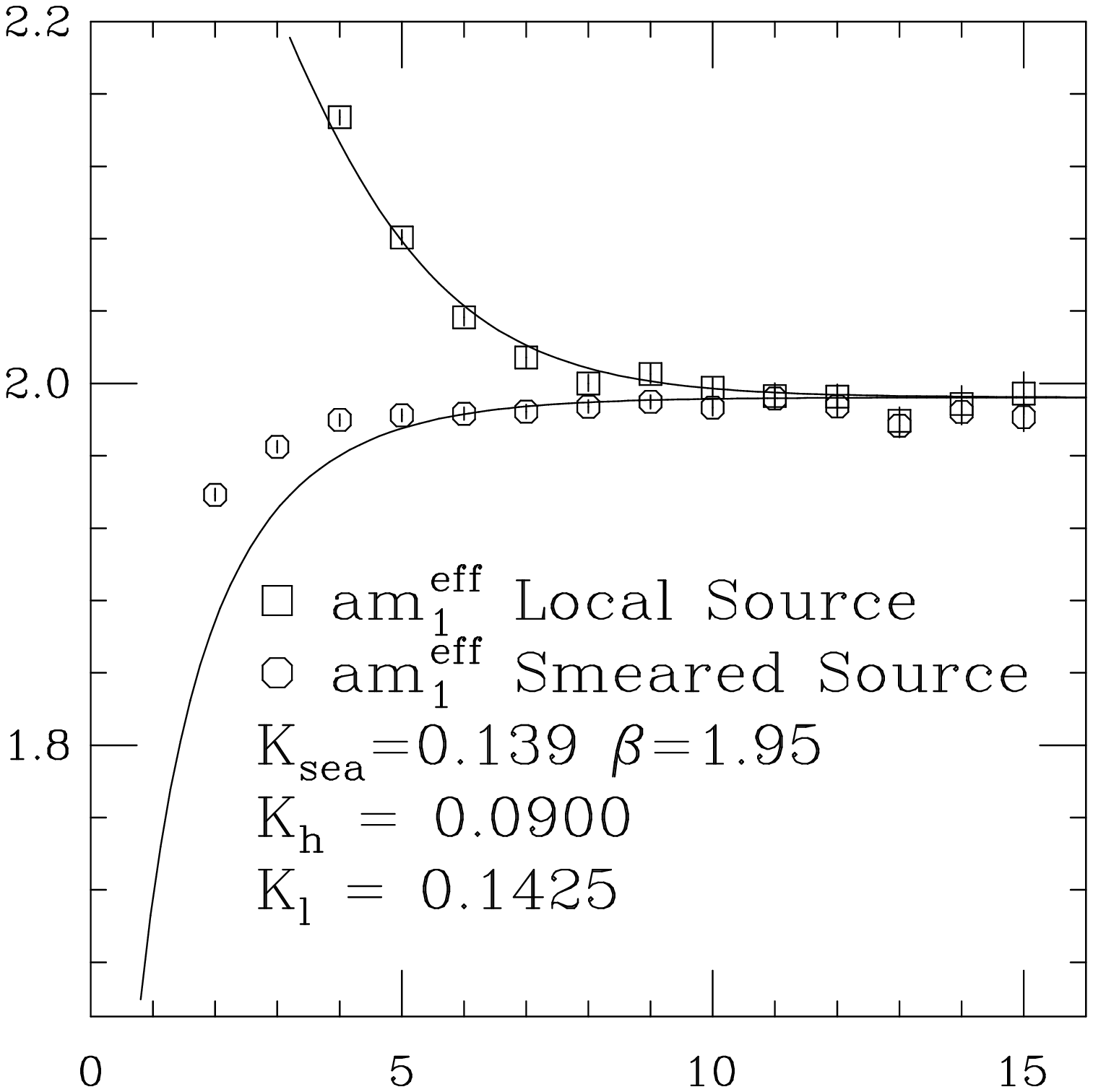}
    \caption{Typical effective mass  plots at $\beta=1.95$ for
      $\nf=2$. 
      The fit range is from 5 to 15.}
  \label{fig:m1eff-B1.95}
  \end{center}
\end{figure}

\begin{figure}
  \begin{center}
    \leavevmode
    \epsfxsize=0.70 \hsize
    \epsffile{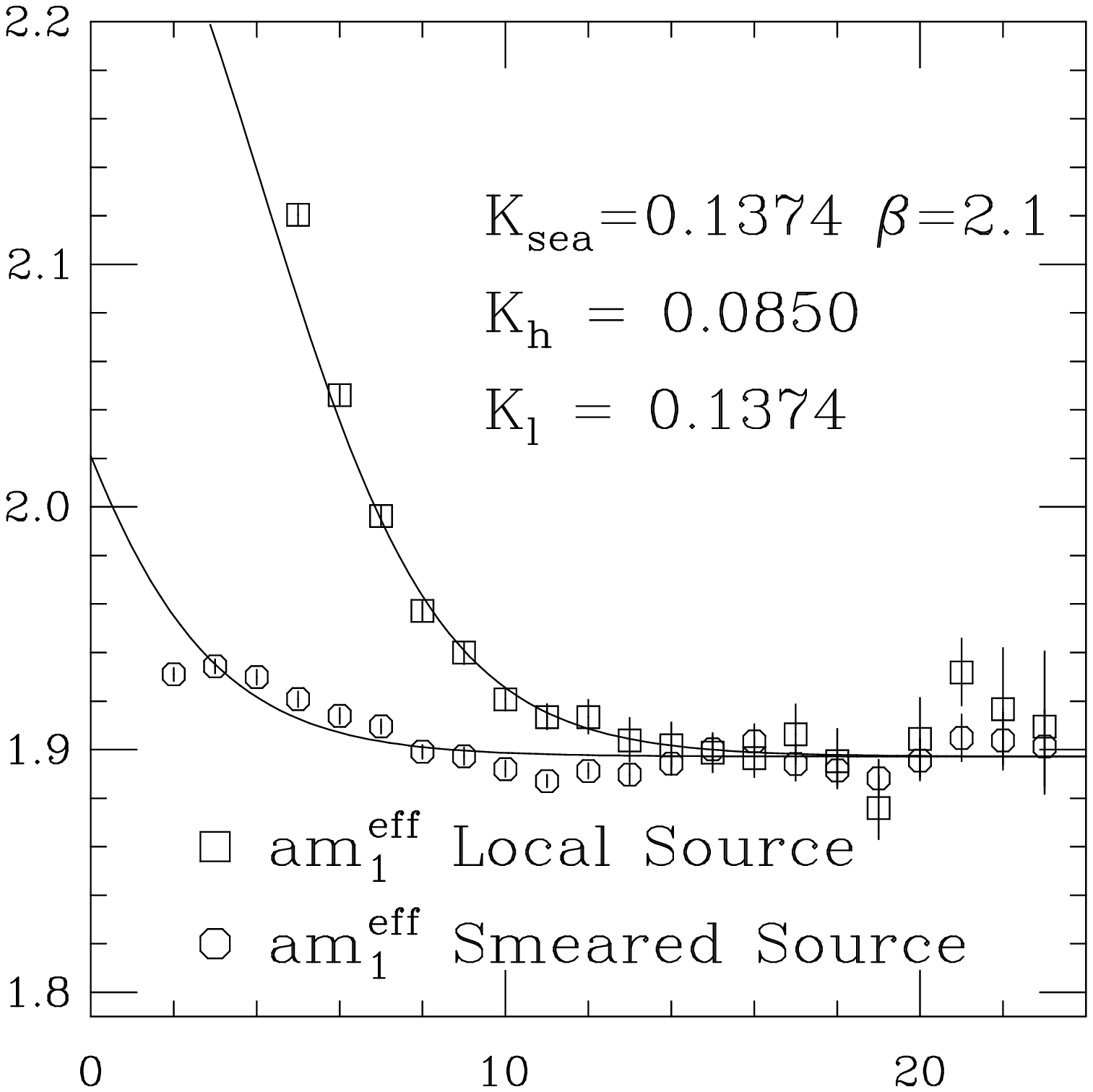}
    \caption{Typical effective mass  plots at $\beta=2.1$ for
      $\nf=2$. 
      The fit range is from 5 to 21.}
  \label{fig:m1eff-B2.1}
  \end{center}
\end{figure}

\begin{figure}
  \begin{center}
    \leavevmode
    \epsfxsize=0.70 \hsize
    \epsffile{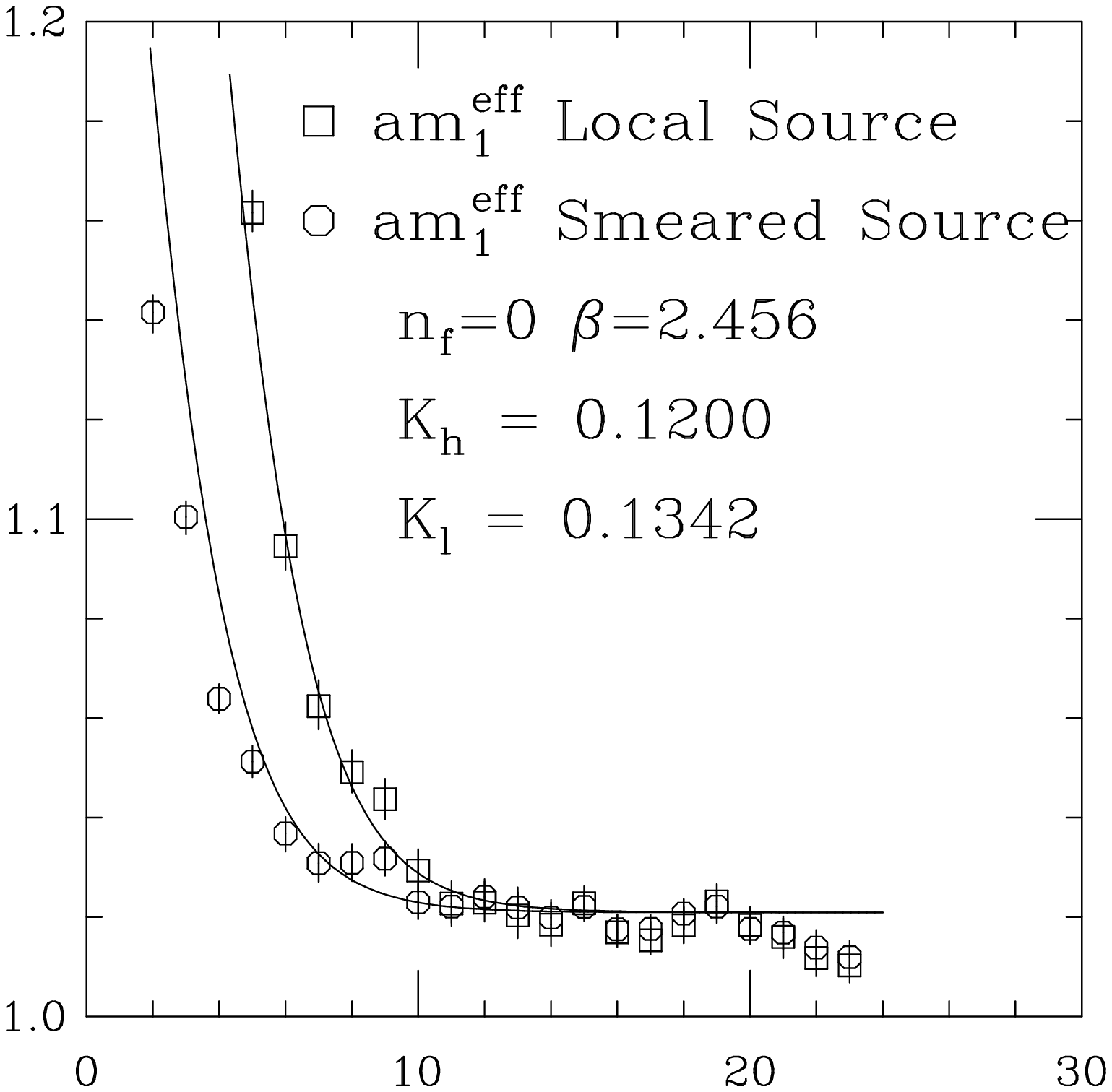}
    \caption{Typical effective mass plots at $\beta=2.456$ for
      $\nf=0$. 
      The fit range is from 5 to 21.}
  \label{m1eff-B2.456}
  \end{center}
\end{figure}

The effective mass plots for the $\langle P P\rangle$ 
correlators, together with fit curves, 
are shown for a typical heavy-light meson mass
in Figures \ref{fig:m1eff-B1.8}--\ref{fig:m1eff-B2.1}
($\nf = 2$ case) and in Figure \ref{m1eff-B2.456}(quenched case).
 
\subsection{Heavy-light decay constant}

\begin{figure}
  \begin{center}
    \leavevmode
    \epsfxsize=0.70 \hsize
    \epsffile{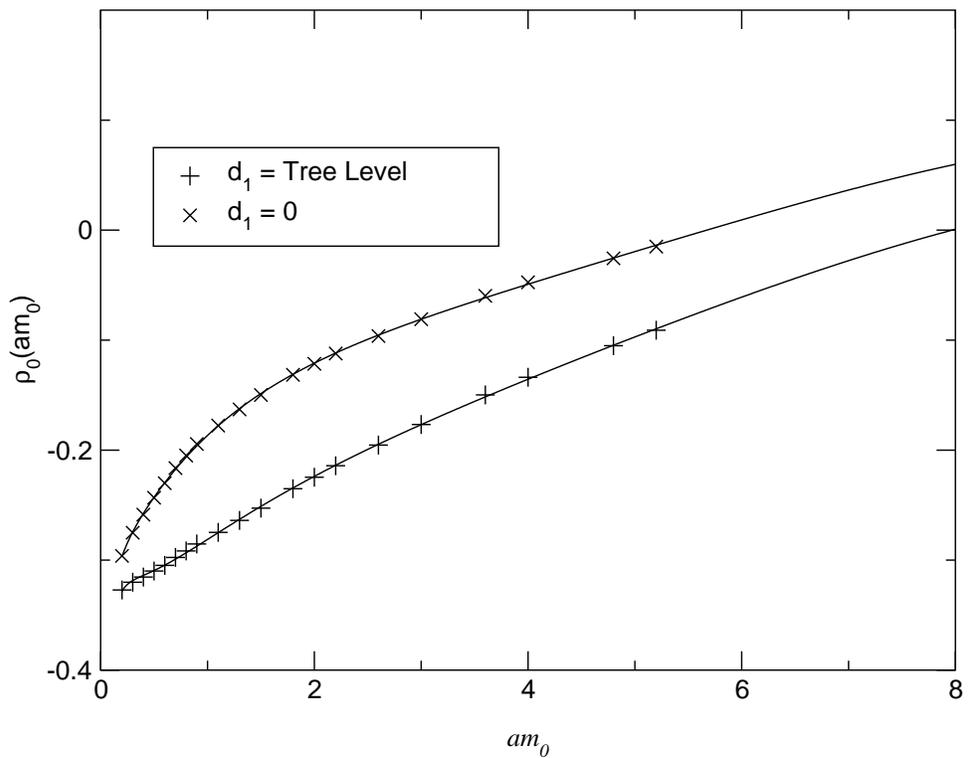}
    \caption{The function $\rho_0(am_0)$ where 
      $Z_A(am_0) = 1 + \alpha_s \rho_0(am_0)$. Pluses show results 
     when the $1/M$ correction to the current is included 
   ({\it i.e.}, $d_1$ takes its tree level value) , and crosses 
   are those without the correction ($d_1=0$).  Solid curves 
   are interpolations.
      }
  \label{rho0}
  \end{center}
\end{figure}

The heavy-light decay constant $f_P$ is obtained through
\begin{equation}
  \label{eq:fsqrtM}
  a^{3/2} (f_P\sqrt{M}) = Z_A \frac{1}{\sqrt{M}}
  \left(
    {\cal Z}_A - a d_1 \delta{\cal Z}_A
  \right).
\end{equation}
In the massless limit, the renormalization constant $Z_A$ was 
previously calculated perturbatively to one-loop order for the RG-improved
action \cite{Aoki:1998ar}.  Here we use a recent extension of this result
to finite heavy quark masses made by K.-I.~Ishikawa {\it et al.}
\cite{Ishikawa}.  
The results can be expressed in the form 
\begin{equation}
  Z_A = 1 + \alpha_s 
  \left[ \rho_A - \frac{1}{\pi} \log (am_0) \right],
\end{equation}
where the one-loop coefficient 
$\rho_0=\rho_A - \frac{1}{\pi} \log (am_0)$ 
is plotted in Figure \ref{rho0} as a function of $am_0$, 
for the cases when $d_1$ takes the tree-level value and 
when it is ignored.

It is well known that perturbative expansions in lattice QCD are 
ill-behaved when one uses the bare coupling constant $g_0^2$, 
and the use of some renormalized coupling defined through
short-distance quantities gives a more convergent expansion \cite{LM}.
Since a two-loop calculation of short-distance
quantities necessary to define an appropriate renormalized coupling 
is not yet available for the RG gauge action, we use the
continuum $\overline{MS}$ coupling as an alternative. 

The one-loop perturbative relation between the bare and $\overline{MS}$
couplings for the RG improved gauge action and the $O(a)$-improved 
Wilson quark action is known as \cite{Aoki:1998ar}
\begin{equation}
  \frac{1}{g^2_{\overline{MS}}(\mu=1/a)} = 
  \frac{\beta}{6} + 0.1000 + 0.0315 \nf.
\end{equation}
The tadpole improvement \cite{LM} may be applied to reduce
the ultraviolet dominated pieces from the perturbative
expansions by reorganizing the above relation as
\begin{equation}
  \frac{1}{g^2_{\overline{MS}}(\mu=1/a)}
    = ( c_0 P - 8 c_1 R ) \frac{\beta}{6} - 0.1006 + 0.0315 \nf,
    \label{eqn:msbardef}
\end{equation}
where 
$P=\langle W_{1\times 1}\rangle$ and 
$R=\langle W_{1\times 2}\rangle$ are the expectation values of 
plaquette and $1\times 2$ rectangle \cite{ref:Iwasaki83}, and 
the one-loop expressions $P = 1 - 0.1402 g^2$ and 
$R = 1 - 0.2689 g^2$ are used to obtain the modified
one-loop coefficient in (\ref{eqn:msbardef}).
The values of $g^2_{\overline{MS}}(\mu=1/a)$ obtained with
this formula are 3.162, 2.812 and 2.562 at $\beta$=1.8, 1.95
and 2.1 respectively for the $\nf=2$ lattices.
The same quantity for the quenched lattices is listed in
Table~\ref{table:lattice-params-nf0}.

\begin{figure}
  \begin{center}
    \leavevmode
    \epsfxsize=0.70 \hsize
    \epsffile{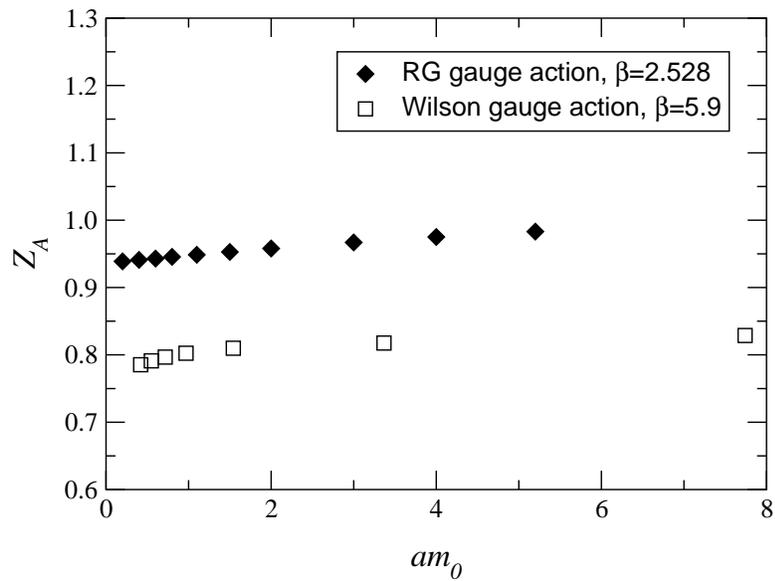}
    \caption{The renormalization constant $Z_A$ as a function
      of $am_0$ for the Wilson (at $\beta$=5.9) and RG (at
      $\beta$=2.528) gauge actions. 
In the case of the RG action, $Z_A$ is computed to specifically
include the $1/M$ correction to the current while in the Wilson 
action it is not.
      The inverse lattice spacing is roughly 1.8 GeV (in the
      quenched approximation) for both cases. 
      }
  \label{Zas}
  \end{center}
\end{figure}

In Figure \ref{Zas} we plot $Z_A$ as a function of the
bare heavy quark mass for the plaquette and RG-improved actions for an
inverse lattice spacing of around 1.8 GeV ($\beta$=5.9 for
the Wilson and $\beta$=2.528 for the RG action in the
quenched approximation). 
In contrast to the large one-loop correction  of order $-$20\% 
for the case of the plaquette gauge action, $Z_A$ is close to unity for
the RG-improved action.

\section{Results}
\label{sec:results}

\subsection{Effect of field rotation to the heavy-light current}
\label{sec:rotation}

We first examine the effect of field rotation (\ref{eqn:field-redef}), 
which is reflected in (\ref{eq:current}) and 
(\ref{eq:fsqrtM}) as a correction proportional to $d_1$.
An order counting suggests the size of the correction
of the order of $a d_1\Lambda_{QCD}$, which is about $d_1\times$15\% 
at $1/a\approx  2$ GeV if $\Lambda_{QCD}=300$ MeV.
Since the tree-level coefficient $d_1$ given by (\ref{eq:d_1}) 
is smaller than 0.1 for any value of the bare quark mass $am_0$, 
the size of the correction is naively estimated to be $O(2\%)$.

\begin{figure}
  \begin{center}
    \leavevmode
    \epsfxsize=0.70 \hsize
    \epsffile{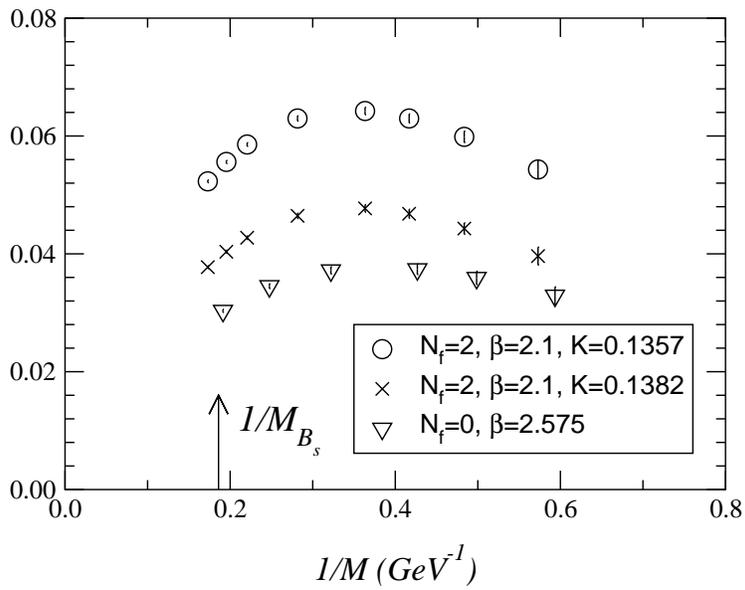}
    \caption{Ratio $f_P^{rotated}/f_P^{unrotated}-1$ 
      of decay constant including the current rotation to the lowest order 
      current to that without the correction
      for $\nf=2$ (circles and crosses) and $\nf=0$
      (triangles). 
      The gauge couplings were picked so that
      the lattice spacing roughly matched with each other.
      } 
  \label{one-one-m-comparison}
  \end{center}
\end{figure}

In Figure \ref{one-one-m-comparison} we plot 
the quantity $f_P^{\mbox{rotated}}/f_P^{\mbox{unrotated}}-1$ 
as a function of the 
meson mass for $\nf=2$ and $\nf=0$, where $f_P^{\mbox{rotated}}$ includes 
the rotation term while it is ignored in $f_P^{\mbox{unrotated}}$. 
Care must be exercised in this comparison to use 
the appropriate renormalization factors $Z_A$ for the rotated and
unrotated currents shown in Figure~\ref{rho0} since the diagram 
originating from the rotation term should be excluded for 
calculating $f_P^{\mbox{unrotated}}$. 
The lattice spacing for $\beta=2.575$ at $\nf=0$ is
approximately equal to the lattice spacing, extrapolated to
the chiral limit, for $\beta=2.1$ at $\nf=2$, which allows
a more relevant comparison of the ratios.
As one can see, the magnitude of correction is of the order
of $3$--$7$\%, which is larger than our expectation and cannot be 
ignored.

The large magnitude of this correction may partly originate from a power 
divergence of the matrix element of the higher dimensional operator
$\delta A$ defined by (\ref{eq:five}), with which the naive order
counting of $O(a\Lambda_{QCD})$ is changed to a size of $O(1)$.
In principle this power divergence should be compensated 
by that in the perturbative matching.
However, at the  one-loop order in the calculation of Ref.~\cite{Ishikawa}, 
the compensation is incomplete.

\subsection{Extrapolation to physical quark masses}
In order to obtain the heavy-light decay constant
$f_P\sqrt{M}$ for the physical mass of $B_{(s)}$ and
$D_{(s)}$ mesons, we fit the data with the following form 
\begin{equation}
  \label{eq:fit_form}
  a^{3/2} \Phi_P =   
  A_0 + A_1 am_q +  A_2 (am_q)^2
  + \frac{1}{aM} \left[ B_0 + B_1 am_q \right]
  + \frac{1}{(aM)^2} C_0, 
\end{equation}
where we define the renormalization group invariant decay constant 
$\Phi_P$ as
\begin{equation}
  a^{3/2} \Phi_P =
  \left(
    \frac{\alpha_s(M)}{\alpha_s(M_B)} 
  \right)^{2/\beta_0}
  a^{3/2} (f_P\sqrt{M}),
\end{equation}
with $\beta_0 = 11-\frac{2}{3}\nf$.
The light quark mass is defined as 
$am_q = \frac{1}{2}\left(\frac{1}{K}-\frac{1}{K_c}\right)$ 
where $K_c$ denotes the value at which pion mass made of sea quarks 
vanishes, and the HQET mass definition (\ref{eq:HQET_mass}) is used for 
the heavy-light meson mass $M$.
The renormalization group factor 
$(\alpha_s(M)/\alpha_s(M_B))^{2/\beta_0}$
is evaluated with a two-loop running coupling coefficient adopting 
$\Lambda_{QCD}$ = 300 MeV for both $N_f=2$ and $N_f=0$. 
We have checked that the resulting decay constants are stable well within 
statistical errors under a variation of $\Lambda_{QCD}$ by a factor two. 

The form (\ref{eq:fit_form}) is a truncated expansion
of the matrix element in $1/aM$ and in $am_q$.
It is possible to include higher order terms; however, the
resulting fit coefficients are statistically not well
determined, and we do not include such terms in our analyses. 

In determining $f_{B_d}$ in the $N_f=2$ case, 
we only employ the matrix elements
where the sea and valence light quark masses are matched.
For $f_{B_s}$ we interpolate, at each sea quark mass, the
matrix element in the valence light quark mass to the
physical strange quark determined  using the partially quenched
analysis \cite{Burkhalter:1998wu,CP-PACS_lightquark,CP-PACS_lighthadron}.
The values of the hopping parameter $K_s$ corresponding to
the strange quark are listed in Tables
\ref{table:lattice-params-nf2} and
\ref{table:lattice-params-nf0},
for the $K$ and $\phi$ meson masses as physical input.
The critical hopping parameter $K_c$ necessary for evaluating 
the light quark mass $m_q$ is also listed in these tables. 
 
The quenched data are analyzed with the same fit ansatz except
that the term $A_2$ is set to zero, as the number of light
quark masses in this case precluded a quadratic fit.
For $\fbs$ and $\fds$, the terms $A_1$ and $B_1$ are also set to
zero, since there is no remaining light quark mass
dependence once the strange quark mass is fixed.

\begin{figure}
  \begin{center}
    \leavevmode
    \epsfxsize=0.70 \hsize
    \epsffile{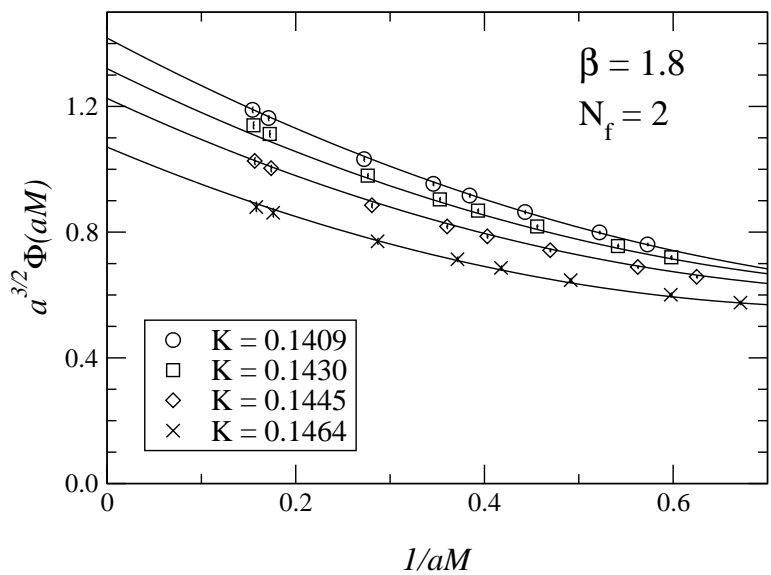}
    \caption{A plot of $\Phi(aM)$ vs $1/aM$ for $\nf=2$ at
      $\beta$=1.8.
      The data for four different sea quark masses are shown,
      and the light valence quark mass is set equal to the sea 
      quark mass.
      }
  \label{phi-fit-nf2-b1.8}
  \end{center}
\end{figure}

\begin{figure}
  \begin{center}
    \leavevmode
    \epsfclipon
    \epsfxsize=0.70 \hsize
    \epsffile{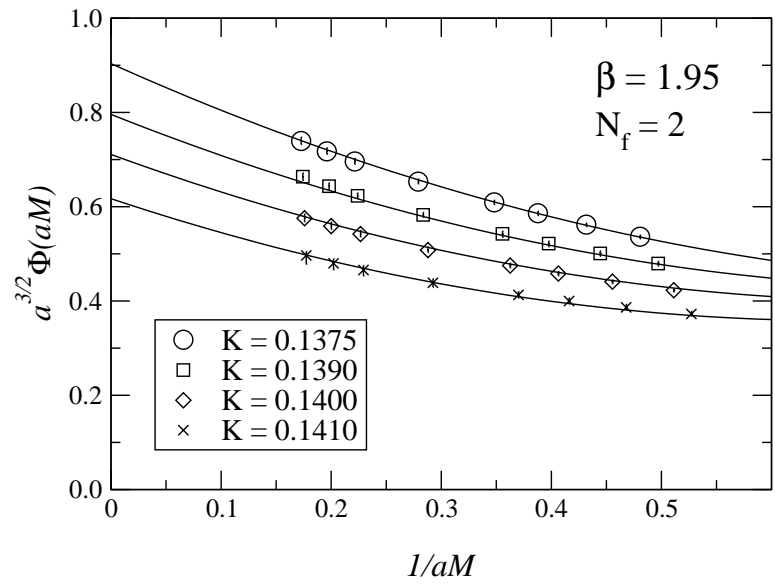}
    \caption{
      Same as Fig.~\ref{phi-fit-nf2-b1.8}, but for
      $\beta=1.95$.
      } 
  \label{phi-fit-nf2-b1.95}
  \end{center}
\end{figure}

\begin{figure}
  \begin{center}
    \leavevmode
    \epsfxsize=0.70 \hsize
    \epsffile{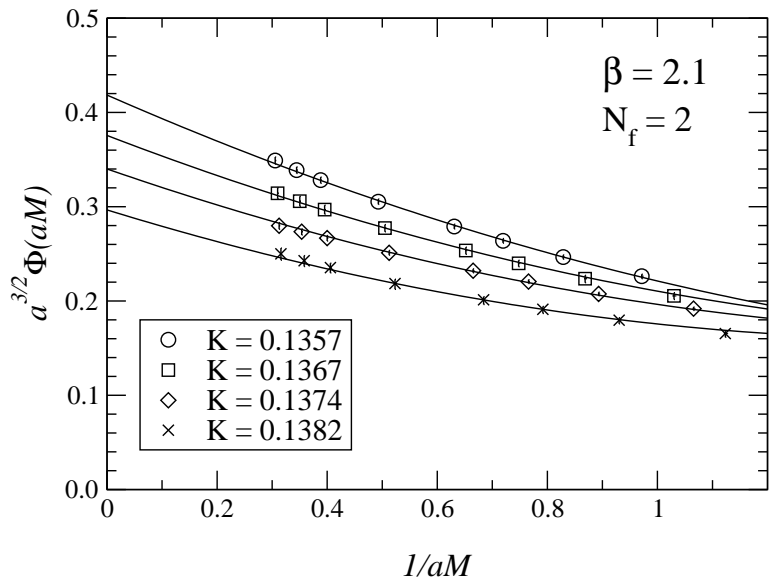}
    \caption{
      Same as Fig.~\ref{phi-fit-nf2-b1.8}, but for
      $\beta=2.1$.
      } 
  \label{phi-fit-nf2-b2.1}
  \end{center}
\end{figure}

\begin{figure}
  \begin{center}
    \leavevmode
    \epsfxsize=0.70 \hsize
    \epsffile{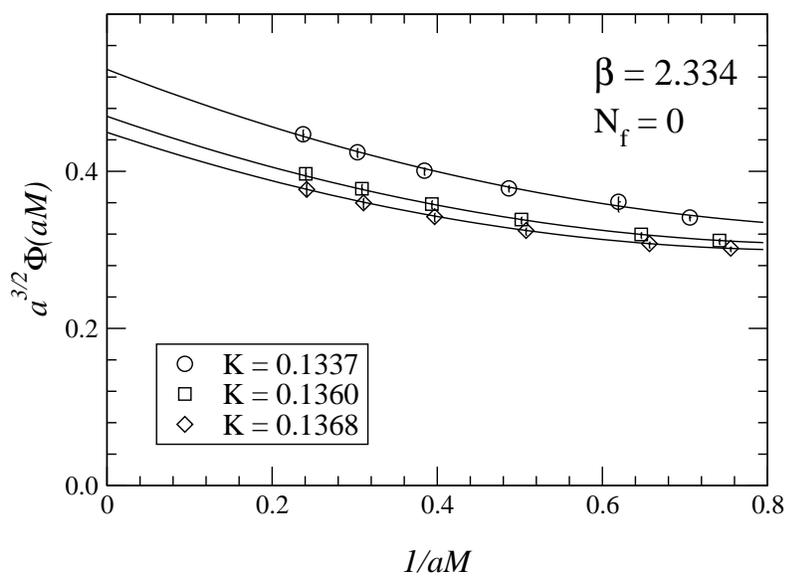}
    \caption{A typical plot of $\Phi(aM)$ vs $1/aM$ for
      $\nf=0$ at $\beta$=2.334.
      }
  \label{phi-fit-nf0}
  \end{center}
\end{figure}

Fits with the form (\ref{eq:fit_form}) are represented in
Figures~\ref{phi-fit-nf2-b1.8}, \ref{phi-fit-nf2-b1.95} and 
\ref{phi-fit-nf2-b2.1} for three lattice spacing for $\nf=2$
and in Figure~\ref{phi-fit-nf0} for $\nf=0$.
Data points and fit curves are plotted as a function of
$1/aM$ for fixed $am_q$, from which one can see that the
ansatz (\ref{eq:fit_form}) represents the data quite well,
except for a few points at $\beta$=1.8.

\begin{figure}
  \begin{center}
    \leavevmode
    \epsfxsize=0.70 \hsize
    \epsffile{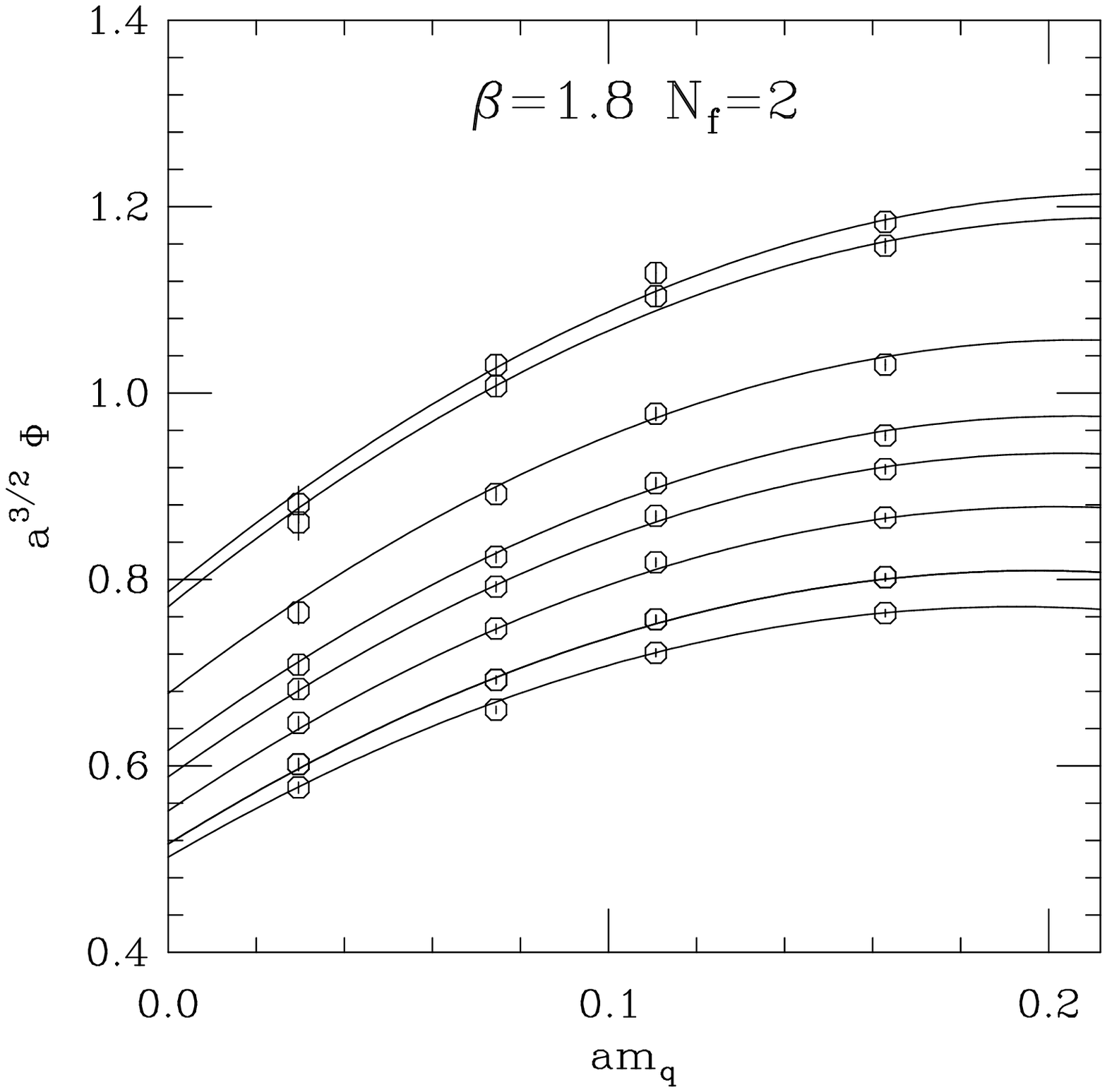}
    \caption{A plot of $\Phi_s$ vs $am_q$ for $\beta=1.80$ and $\nf=2$.}
  \label{phi-vs-mpisq-b18-nf2}
  \end{center}
\end{figure}

\begin{figure}
  \begin{center}
    \leavevmode
    \epsfxsize=0.70 \hsize
    \epsffile{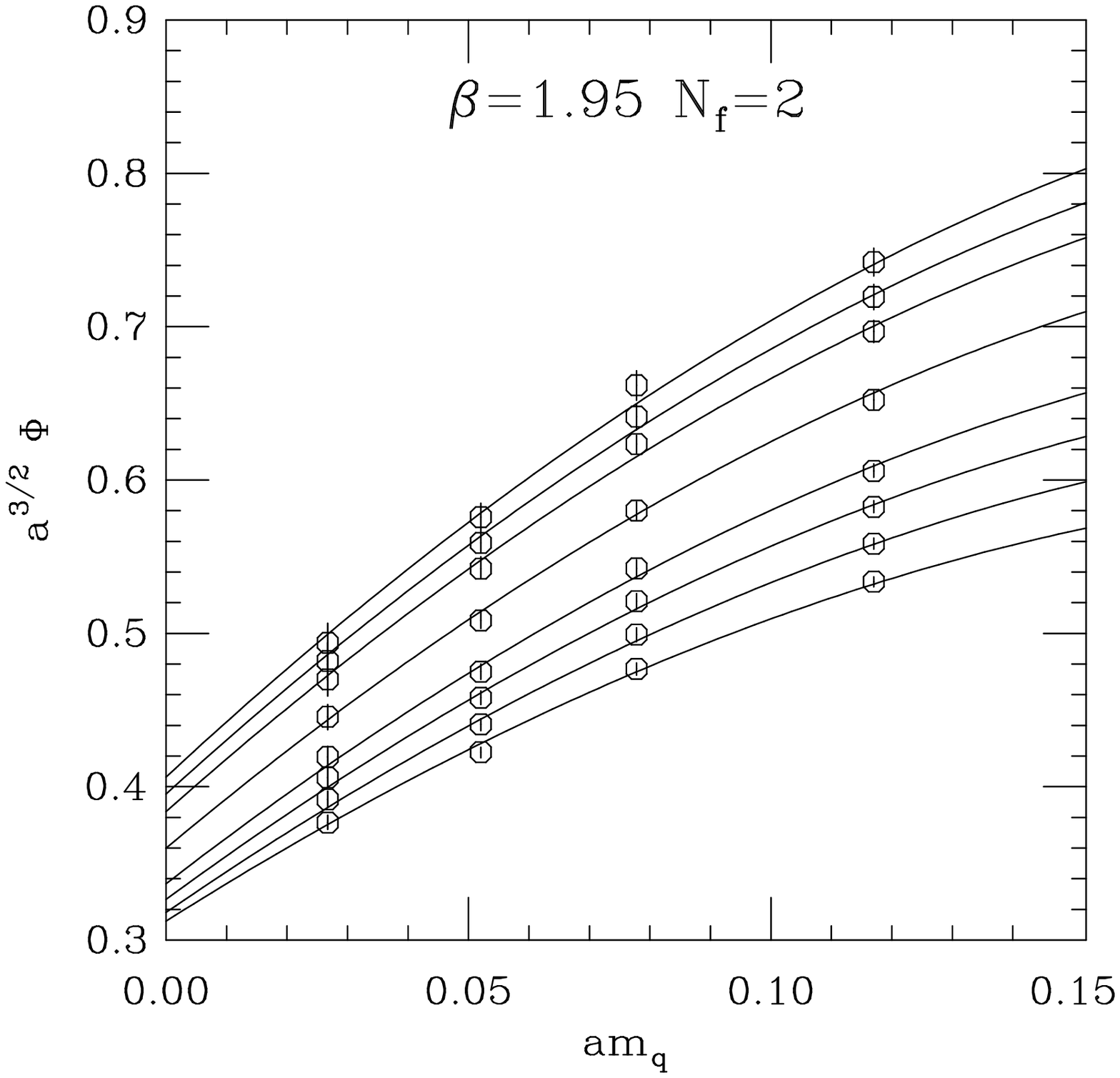}
    \caption{A plot of $\Phi_s$ vs $am_q$ for $\beta=1.95$ and $\nf=2$.}
  \label{phi-vs-mpisq-b195-nf2}
  \end{center}
\end{figure}

\begin{figure}
  \begin{center}
    \leavevmode
    \epsfxsize=0.70 \hsize
    \epsffile{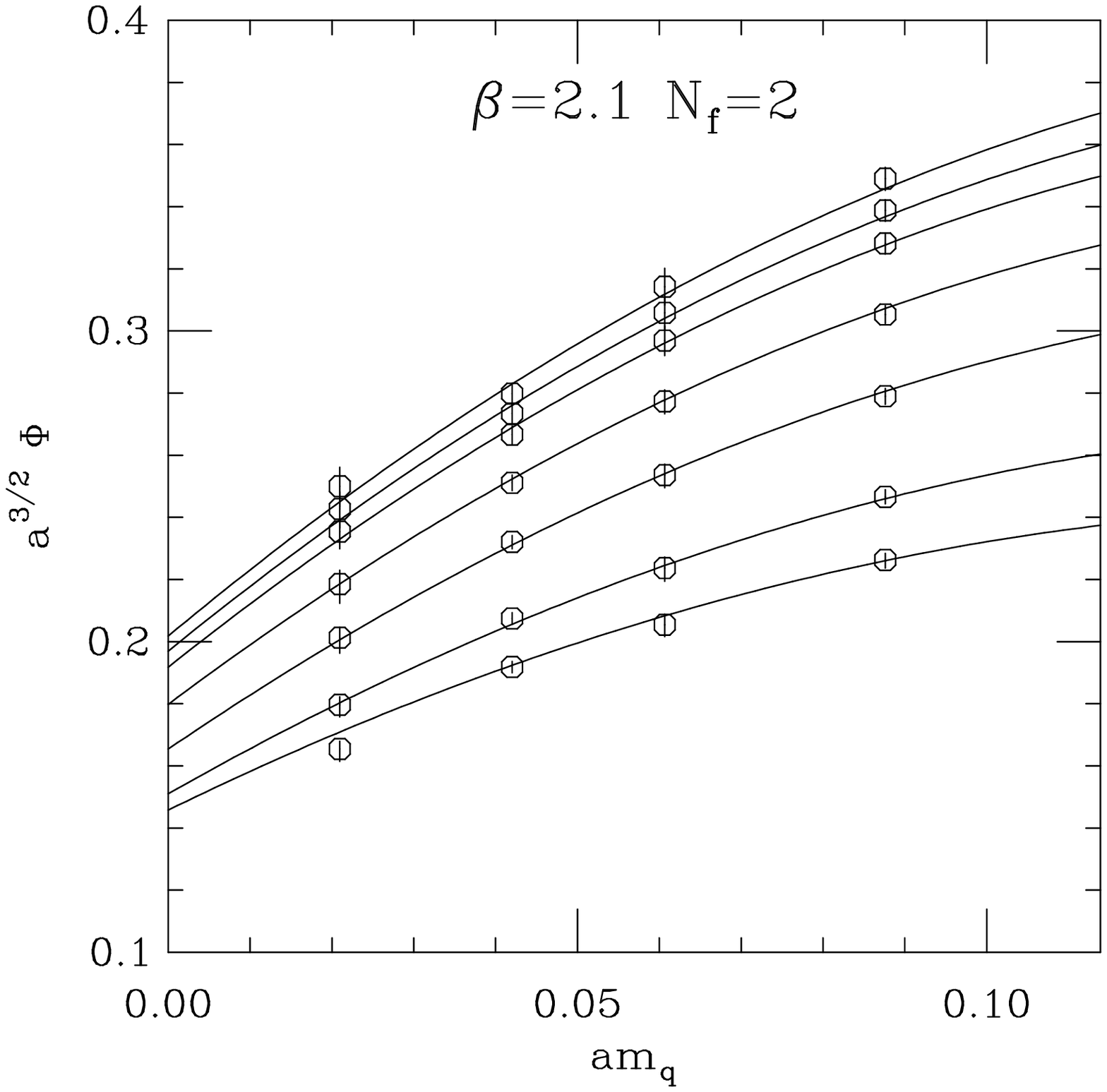}
    \caption{A plot of $\Phi_s$ vs $am_q$ for $\beta=2.10$ and $\nf=2$.}
  \label{phi-vs-mpisq-b21-nf2}
  \end{center}
\end{figure}

\begin{figure}
  \begin{center}
    \leavevmode
    \epsfxsize=0.70 \hsize
    \epsffile{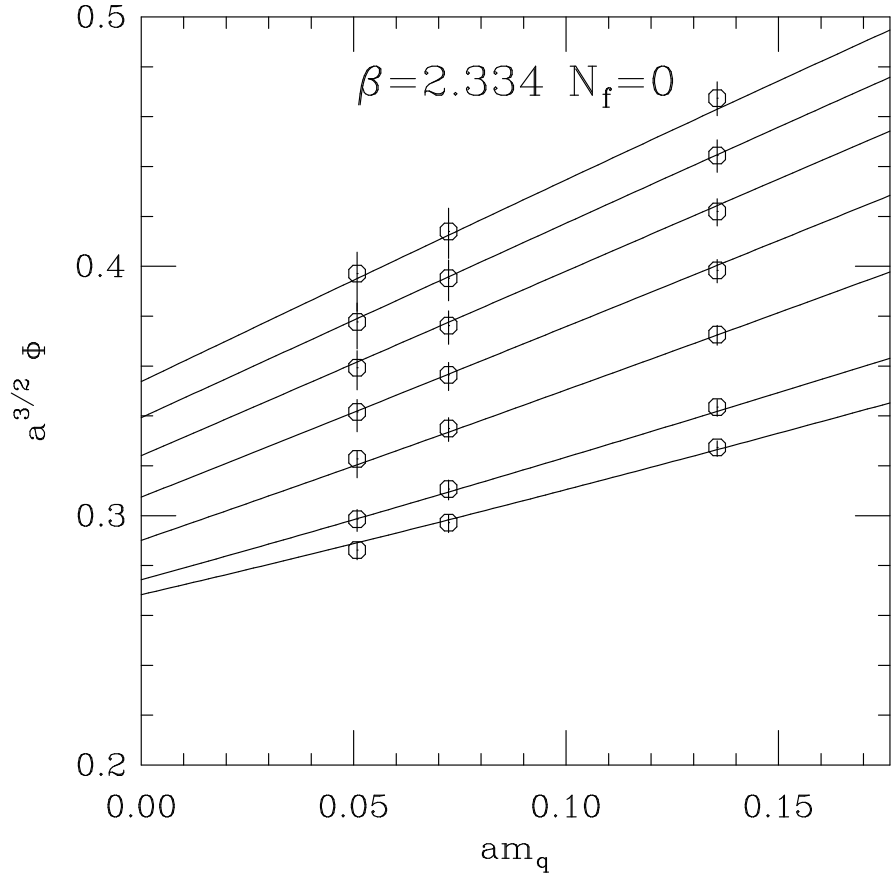}
    \caption{A typical plot of $\Phi$ vs $am_q$ for $\nf=0$.
      For $\nf=0$, $am_q$ is the bare quark mass.}
  \label{phi-vs-mpisq-nf0}
  \end{center}
\end{figure}

It is also illustrative to plot the data at fixed heavy
quark masses as a function of $am_q$, which is shown in 
Figures~\ref{phi-vs-mpisq-b18-nf2}--\ref{phi-vs-mpisq-b21-nf2}
for $\nf=2$ and 
Figure \ref{phi-vs-mpisq-nf0} for $\nf=0$.
Since the results are given for fixed $K_h$, 
the heavy hopping parameter, we interpolated 
the curves for $aM$ as a function of $K_h$ and $am_q$ and  hence
reexpressed the coefficients
of (\ref{eq:fit_form}) as a function of $K_h$ and $am_q$. 
For $\nf=2$ we find clear curvature, which motivated us to
introduce the term $A_2 (am_q)^2$ in (\ref{eq:fit_form}).
We, then, find good agreement of the fits to the data
points. 
The fit parameters $A_i$, $B_i$ and $C_0$
are summarized in
Tables~\ref{table:chiral-hqet-params-nf0}--\ref{table:strange-phi-hqet-params-nf2}
for each set of configurations.

The $B$ and $D$ meson decay constants in 
physical units are obtained from (\ref{eq:fit_form}) with
their physical meson masses as input, and the numbers are
summarized in Tables \ref{table:final-results-B} and
\ref{table:final-results-D} respectively. 
The lattice scale is set using the $\rho$ meson mass.

\subsection{Discretization effect}

\begin{figure}
  \begin{center}
    \leavevmode
    \epsfclipon 
    \epsfxsize=0.70 \hsize
    \epsffile{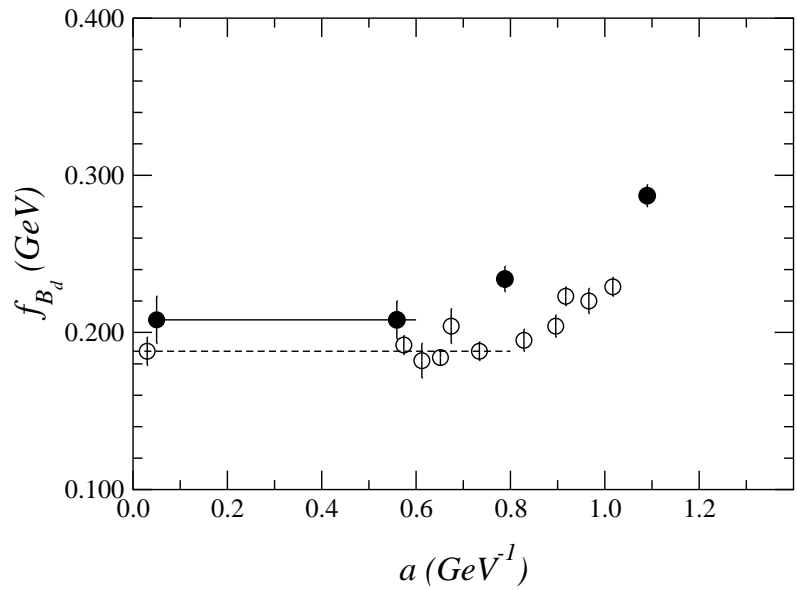}
    \caption{
      $\fb$ for $\nf=2$ (filled circles) and $\nf=0$
      (open circles) as a function of lattice spacing $a$.
      The error bar for the data points represents the
      statistical errors only, while those in the continuum
      limit ($a$=0) are the systematic and statistical errors
      added in quadrature.
      }
  \label{fB-vs-rho}
  \end{center}
\end{figure}

\begin{figure}
  \begin{center}
    \leavevmode
    \epsfclipon 
    \epsfxsize=0.70 \hsize
    \epsffile{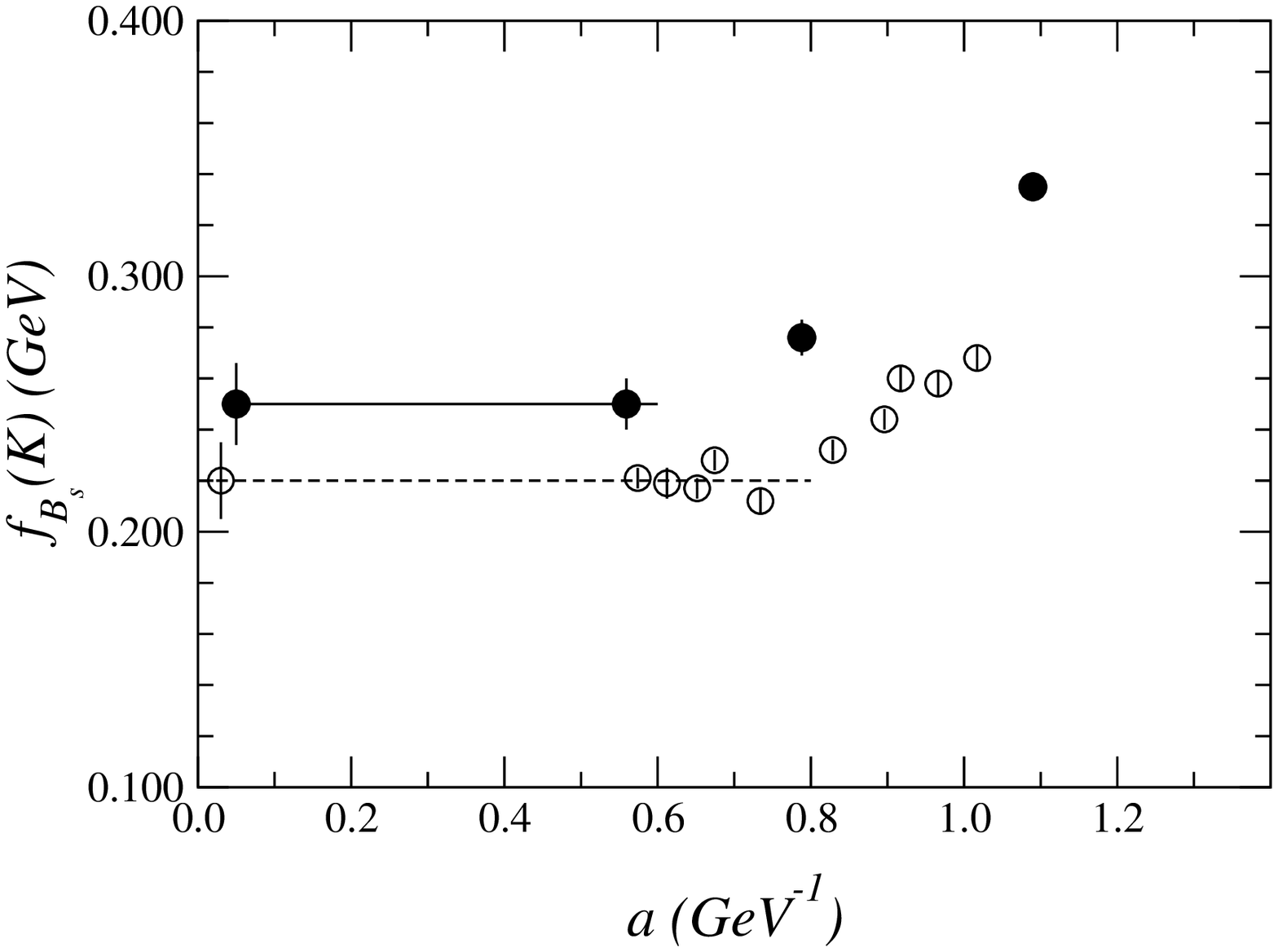}
    \caption{
      Same as Figure~\ref{fB-vs-rho}, but for $f_{B_s}(K)$.
      }
  \label{fBs-vs-rho}
  \end{center}
\end{figure}

\begin{figure}
  \begin{center}
    \leavevmode
    \epsfclipon 
    \epsfxsize=0.70 \hsize
    \epsffile{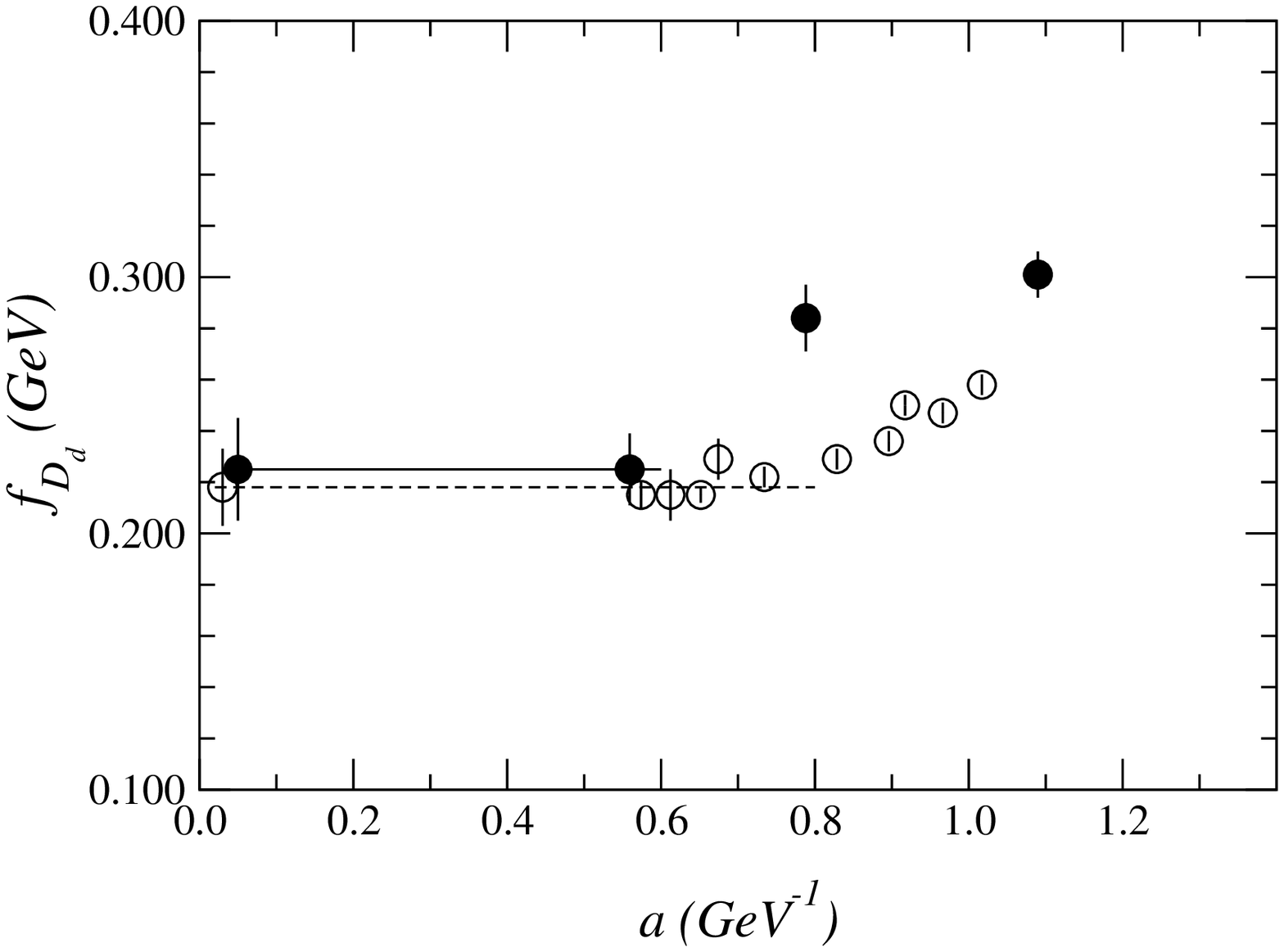}
    \caption{
      Same as Figure~\ref{fB-vs-rho}, but for $f_{D_d}$.
      }
  \label{fD-vs-rho}
  \end{center}
\end{figure}

\begin{figure}
  \begin{center}
    \leavevmode
    \epsfclipon 
    \epsfxsize=0.70 \hsize
    \epsffile{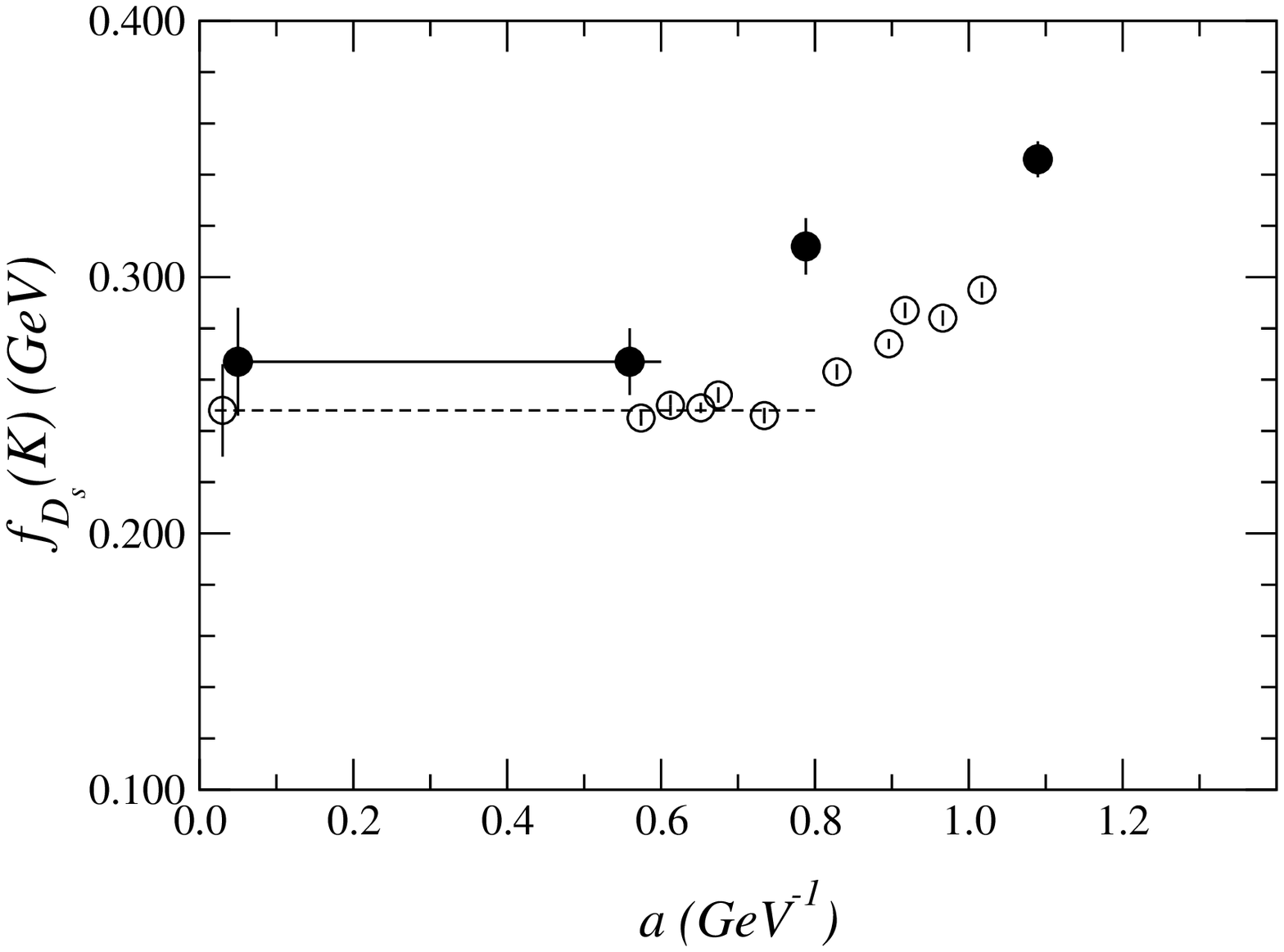}
    \caption{
      Same as Figure~\ref{fB-vs-rho}, but for $f_{D_s}(K)$.
      }
  \label{fDs-vs-rho}
  \end{center}
\end{figure}

The decay constants are plotted as a function of $a$ in Figures
\ref{fB-vs-rho}($f_{B_d}$), \ref{fBs-vs-rho}($f_{B_s}$), 
\ref{fD-vs-rho}($f_{D_d}$) and \ref{fDs-vs-rho}($f_{D_s}$).
For $f_{B_s}$ (Figure \ref{fBs-vs-rho}) and $f_{D_s}$
(Figure \ref{fDs-vs-rho}), we use the mass of $K$ to
define the strange quark mass (as a short hand, we will
refer to this as $f_{B_s}(K)$ and $f_{D_s}(K)$).  

For the quenched data ($\nf$ = 0), where ten data points are
available, we observe a linear decrease as the lattice spacing 
decreases from $a\approx 1$~GeV$^{-1}$ to $a \approx $ 0.8 GeV$^{-1}$, 
followed by an almost constant behavior within statistical fluctuations
below $a \approx $ 0.8 GeV$^{-1}$. 
We therefore fit the five data points for
$a \lsim \, 0.8\, \GeV^{-1}$ as shown in the figures, and take this 
as our best estimate for the quenched value of the decay constant.

There is no evidence that $N_f=2$ data become independent of the lattice
spacing. So we are not able to safely estimate the $B$ meson decay constant
from our data. We may discuss, however, our results in the following way.
The slope of the decrease of $f_B$ from $\beta=1.8$ to 1.95 quite 
resembles that for $N_f=0$ for stronger couplings, while the
decrease from $\beta=1.95$ to 2.1 is somewhat reduced.
If we suppose that the $N_f=2$ data behave in a way similar to those for 
$N_f=0$,
the $N_f=2$ data would have already reached the asymptotic flattening at 
around $a\approx 0.7 \mbox{GeV}^{-1}$ and the data at $\beta=2.1$ may be 
taken as a 
continue value.  Since we cannot do better with the present data, we
provisionally take the point at $\beta=2.1$ as the continuum value, 
allowing for
the possibility that the true value may be somewhat smaller than our estimate.
From the shape of the beta dependence of the $N_f=2$ data and 
their error bars, 
however, we may still safely conclude that the continum value does not 
undershoot that of $N_f=0$.

The extraction of the continuum limit for the $D$ meson decay 
constant is more subtle, since 
we see a larger drop from $\beta=1.95$ to $\beta=2.1$ rather 
than $\beta=1.8$ to
1.95. While we take the data at $\beta=2.1$ as our provisional estimate for
$f_D$, there is a possibility that the true value is smaller, although we
expect that it is equal to or larger than quenched result from the
experience with the $B$ decay constant. Our safe conclusion is that the
dynamical effect for $D$ mesons is apprecaibly smaller than that for $B$ 
mesons.   

\begin{figure}
  \begin{center}
    \leavevmode
    \epsfclipon 
    \epsfxsize=0.70 \hsize
    \epsffile{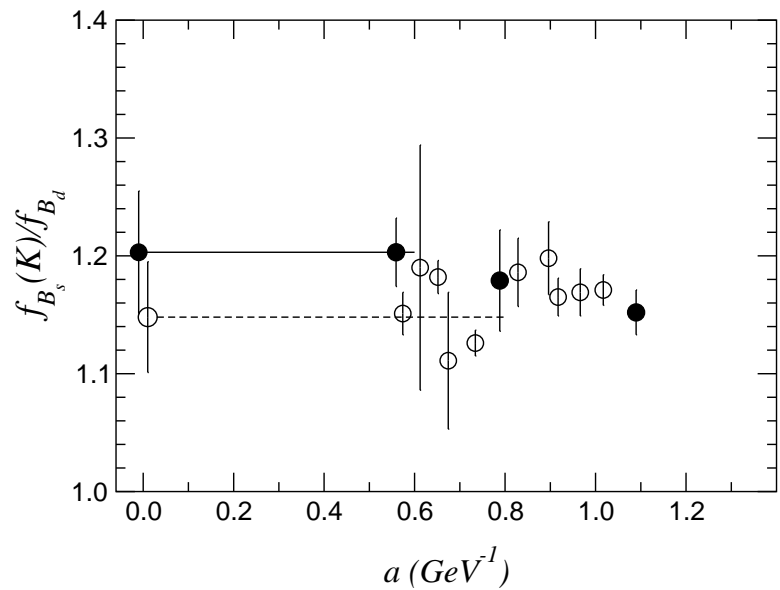}
    \caption{
      A comparison of the ratio $\fbs(K)/\fb$ for $\nf=0$ and
      $\nf=2$.
      The error bars of the continuum limit results are the
      systematic and statistical errors added in quadrature.
      }
  \label{fig:B-ratio}
  \end{center}
\end{figure}

\begin{figure}
  \begin{center}
    \leavevmode
    \epsfclipon 
    \epsfxsize=0.70 \hsize
    \epsffile{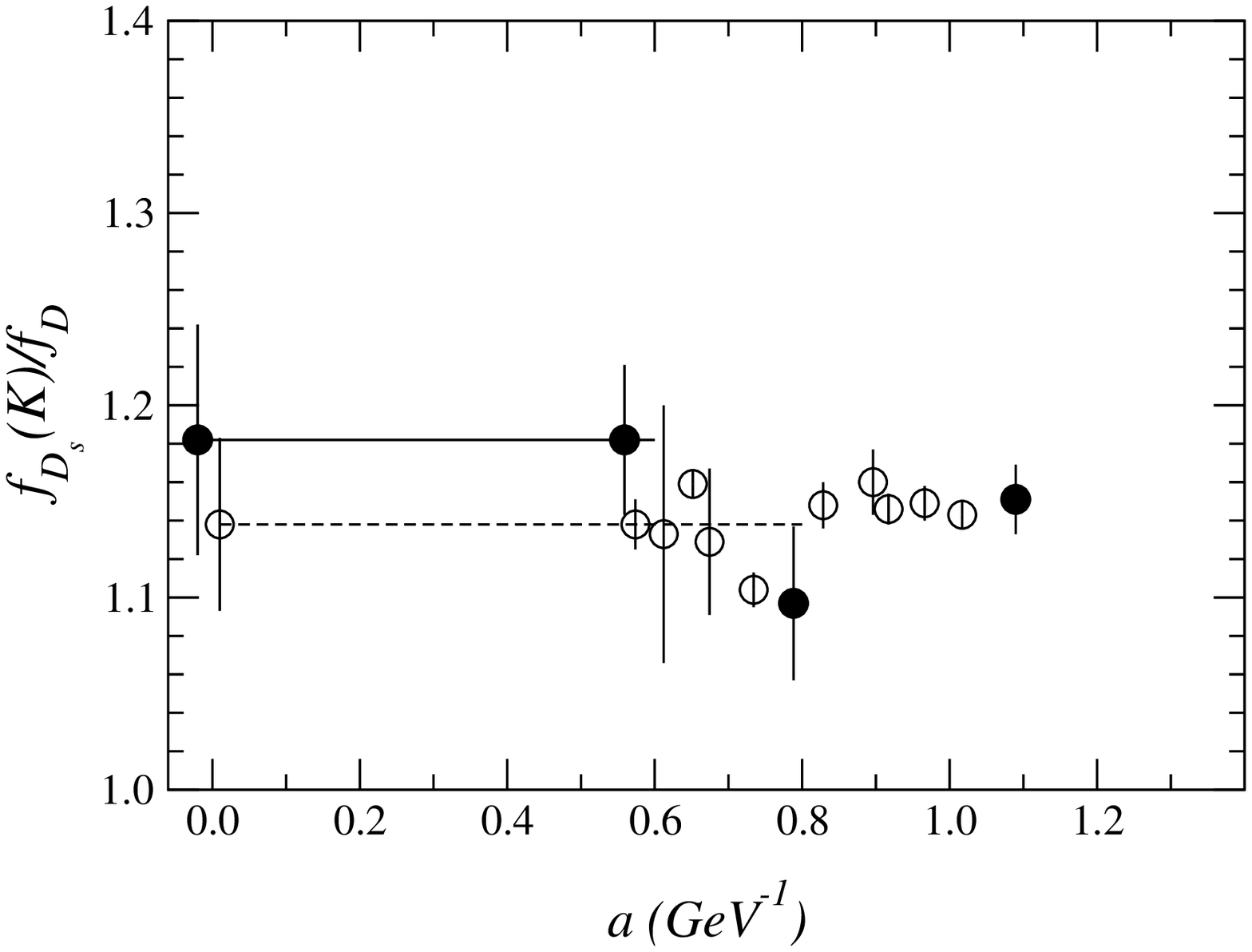}
    \caption{
      Same as Figure~\ref{fig:B-ratio}, but for $\fds(K)/\fd$.
      }
  \label{fig:D-ratio}
  \end{center}
\end{figure}

We employ the same strategy as above for estimating the ratios
$\fBs/\fBd$ and $\fDs/\fDd$ 
as shown in Figures \ref{fig:B-ratio} and \ref{fig:D-ratio}. 

\subsection{Systematic errors}

We now examine the issue of systemtic errors in our results 
for the decay constants. For this purpose we list the possible
leading order errors and estimate their magnitude using naive 
power counting. 
   
Generically these errors appear in three forms.  
First, we use tree-level mean-field
estimates of the coefficients in the actions and currents
and hence there will be radiative corrections, which are 
proportional to some power of $\alpha_s(\mu)$.
Since the dominant part of the radiative corrections comes
from a short distance region in the lattice four-momentum
integral, we assume the scale $\mu$ to be $1/a$. 
Secondly, discretization effects in the lagrangians will be
of the order of  $(a|{\mathbf{p}}|)^{n}$, where $n$ is an
integer and  $\mathbf{p}$ is some soft momentum scale that
characterizes the spatial momentum of the system. 
We take these soft modes to be of the order of
$\Lambda_{QCD}$. 
Finally, there are power corrections to the heavy quark
effective Hamiltonian, which are of the order of some power of
$\Lambda_{QCD}/M$.

In detail we expect the following corrections in our case.  
(i) Gluon and light quark actions: 
  For the RG-improved gauge action, the leading discretization error
  is of the same order as the plaquette gauge action, which is 
  $(a \Lambda_{QCD})^2$. 
  For the $O(a)$-improved Wilson quark action for light quarks, 
  the leading error is of $O(\alpha_s a\Lambda_{QCD})$ since 
  the coefficient $c_{SW}$ is tuned at one-loop level only by 
  the mean field improvement.
(ii) Heavy quark action:
  For the $O(a)$-improved Wilson quark action, the leading
  error appears in the $1/M^2$ term at the tree level in the effective
  Hamiltonain, which is a source of systematic error of 
  $O((\Lambda_{QCD}/M)^2)$.
  An additional error comes from the radiative correction 
  that changes the relation between $M_2$ and $M_B$, and
  yields an uncertainty of $O(\alpha_s \Lambda_{QCD}/M)$.
(iii) Current corrections:
  The renormalization coefficient $Z_A$ is computed only to one-loop
  accuracy, hence higher order uncertainties are of the order of $\alpha_s^2$. 
  Other corrections to the current are present, but these are of the same 
  order as those in the heavy quark effective Hamiltonian. 

The size of these corrections are estimated in Tables
\ref{table:lattice-sys-nf2} and \ref{table:lattice-sys-nf0}
for the $B$ and $D$ mesons.  Numerical values are evaluated 
adopting the $\overline{MS}$ coupling at the scale $\mu=1/a$ 
defined by (\ref{eqn:msbardef}) and $\Lambda_{QCD}=300$~MeV, 
and substituting in $M$ the physical $B$ or $D$ meson mass. 
In the case of $\nf$=0, we only choose three representative 
$\beta$ values as the variation of the errors is so mild across 
the available range of lattice spacings.
The total uncertainty is estimated by adding all
individual sources in quadrature.

\begin{figure}
  \begin{center}
    \leavevmode 
    \epsfclipon 
    \epsfxsize=0.70 \hsize
    \epsffile{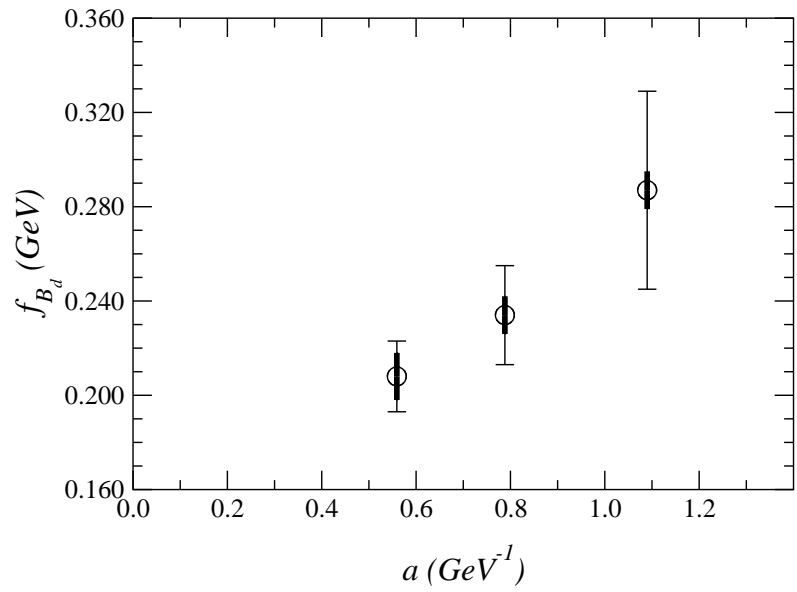}
    \caption{
      $f_{B_d}$ with combined statistical and systematic
      errors for $\nf=2$.
      The statistical error is shown by thick error bars,
      while thin lines represent the total error for which
      the statistical and estimated systematic errors are
      added in quadrature.
      }
  \label{fB-systematic}
  \end{center}
\end{figure}

In Figure \ref{fB-systematic} we replot $f_{B_d}$ for $N_f=2$ 
as a function of lattice spacing.  
The statistical error is shown by thick error bars, 
while thin lines represent the total error for which the statistical 
and estimated systematic errors are added in quadrature. 
The systematic error is the smallest at the finest lattice spacing
($\beta=2.1$) as one can see in Table \ref{table:lattice-sys-nf2}. 
This confirms our expectation that the result from this $\beta$ value 
provides the best estimate for $f_{B_d}$ in the continuum limit.  
It is also important to note that the the data at coarser lattice spacings,
especially that at $\beta$=1.95, are 
consistent with the result at $\beta$=2.1, if we take the systematic error 
into account. This bolsters our confidence in the estimate of the systematic 
error. 

\subsection{Continuum estimate}

In Figures \ref{fB-vs-rho}--\ref{fDs-vs-rho} we plot our final results, 
including the estimated total error,  
for the continuum value of the heavy-light decay constants at $a=0$. 
For the $\nf=2$ calculation with dynamical quarks, 
the central value is taken from the data at the finest lattice spacing
($\beta$=2.1), and the total error shown is obtained by 
quadratically adding the statistical and systematic errors. 
Numerically, we find for $\nf=2$,
\begin{eqnarray}
  \fBd^{\nf=2} &=& 208(10)(11) \; \MeV, \\
  \fBs^{\nf=2} &=& 250(10)(13)(^{+8}_{-0}) \; \MeV, \\
  \left(\frac{\fBs}{\fBd}\right)^{\nf=2} &=& 
  1.203(29)(43)(^{+38}_{- 0}), \\
  \fDd^{\nf=2} &=& 225(14)(14) \; \MeV, \\
  \fDs^{\nf=2} &=& 267(13)(17)(^{+10}_{-0}) \; \MeV, \\
  \left(\frac{\fDs}{\fDd}\right)^{\nf=2} &=& 
  1.182(39)(46)(^{+41}_{- 0}). 
\end{eqnarray}
The first error is statistical, and the second error is the 
cumulative systematic error outlined above. 
For the ratios, as ambiguities due to the renormalization
coefficient are eliminated, only the effect of the gluonic
and light quark errors are included.
In the case of those quantities involving the strange quark,
the central value was taken from the strange quark mass
defined from $m_K$, while a systematic error was estimated
from mass of the $\phi$.
If instead of adding the systematic errors quadratically, 
we added them linearly, the 
final results, taking $\fb$ as an example, 
would be $\fb$ = 208(10)(19) MeV.
It is encouraging that our prediction for $f_{D_s}$ with
$\nf=2$ is consistent with the recent experiments
285$\pm$20$\pm$40 MeV (ALEPH \cite{ALEPH}) and
280$\pm$19$\pm$44 MeV (CLEO \cite{CLEO}).

We also quote the results for the quenched case $\nf=0$, for
which we employ a constant fit to the five data points in the region 
$a \lsim \, 0.8\,\GeV^{-1}$ corresponding to $\beta$ = 2.575--2.416.
The estimated systematic error varies only slightly in this
region,  and we find for $\nf=0$, 
\begin{eqnarray}
  \fBd^{\nf=0} &=& 188(3)(9) \; \MeV, \\
  \fBs^{\nf=0} &=& 220(2)(15)(^{+8}_{-0}) \; \MeV, \\
  \left(\frac{\fBs}{\fBd}\right)^{\nf=0} &=& 
  1.148(8)(46)(^{+39}_{- 0}), \\
  \fDd^{\nf=0} &=& 218(2)(15) \; \MeV, \\
  \fDs^{\nf=0} &=& 250(1)(18)(^{+6}_{-0}) \; \MeV, \\
  \left(\frac{\fDs}{\fDd}\right)^{\nf=0} &=& 
  1.138(5)(45)(^{+29}_{- 0}), \\
\end{eqnarray}
where the systematic errors are assigned with the same
strategy as for the case of $\nf=2$.

These quenched decay constants lie at the 
upper end when compared with those from previous quenched lattice 
calculations, whose recent summary is 
$\fBd$ = 170(20)~MeV, $\fBs$ = 195(20)~MeV
\cite{Hashimoto:1999bk}, and 
$\fDd$ = 200(20)~MeV, $\fDs$ = 220($^{+25}_{-20}$)~MeV
\cite{Draper_lat98}.
One of the differences in this work from the previous ones is
in the choice of the gauge action, with which the
perturbative matching factor is significantly different.  
Hence two-loop uncertainties of order 5\% or so should be 
allowed for in the comparison.   
Another difference is the treatment of the $O(1/M)$ correction 
arising from the field rotation (\ref{eqn:field-redef}),  
which is included in our calculation but not in the previous studies. 
As we discussed in Sec.~\ref{sec:rotation} this term 
increases the value of decay constants.
If we eliminate the $1/M$ correction in our calculation, we find 
$\fBd^{\nf=0}=182(3)(9)\;\MeV$
$\fBs^{\nf=0}=213(2)(14)\;\MeV$,
$\fDd^{\nf=0}=208(2)(14) \; \MeV$ and
$\fDs^{\nf=0}=238(1)(17) \; \MeV$, {\it i.e.,} a 3\% effect for $B$
and a 6\% effect for $D$.   
Combined with uncertainties from the matching factors dicussed above, 
we consider that our quenched results with the RG-improved action are 
consistent with the previous data obtained with the 
plaquette gauge action.

\subsection{Quenching effects}

In order to elucidate the effect of introducing sea quarks
it is instructive to take the ratio of the results for $\nf=2$ and $\nf=0$, 
for which we find :
\begin{eqnarray}
\frac{\fb^{\nf=2}}{\fb^{\nf=0}} &=& 1.11(6) \;\; , \\
\frac{\fbs^{\nf=2}}{\fbs^{\nf=0}} &=& 1.14(5) \;\; , \\
\left(\frac{\fbs}{\fb}\right)^{\nf=2} {\Bigg /} \left(\frac{\fbs}{\fb}\right)^{\nf=0} 
& = & 1.05(3) \;\; , \\
\frac{\fd^{\nf=2}}{\fd^{\nf=0}} &=& 1.03(6) \;\; , \\
\frac{\fds^{\nf=2}}{\fds^{\nf=0}} &=& 1.07(5) \;\; , \\
\left(\frac{\fds}{\fd}\right)^{\nf=2}  {\Bigg /}\left(\frac{\fds}{\fd}\right)^{\nf=0} 
& = & 1.04(3) \;\; . 
\end{eqnarray}
 
The errors quoted above are statistical only since we expect 
the systematic error to largely cancel in the ratio.
We observe that the magnitude increases by $10$--$15$\% 
for the $B$ meson decay constants 
when two flavours of  dynamical quarks are introduced, 
which has statistical significance of 2 to 3 standard deviations. 
For the $D$ meson decay constant, on the other hand, 
the observed increase is only $3$--$7$\%, 
and the effect is statistically less significant.
For the ratio of decay constants we find only a small 
change from $\nf=0$ to $\nf=2$.

An increase of $B$ meson 
decay constants in the presence of dynamical sea quarks 
has already been suggested in Refs.~\cite{Bernard:1998xi,Collins:1999ff}. 
Our results also show this trend, providing further evidence that 
the upward shift is real.

\section{Conclusions}
\label{sec:conclusion}

In this paper we have presented a calculation of the heavy-light axial 
decay constants $\fb$, $\fbs$, $\fd$, $\fds$ and their ratios
in lattice QCD with two degenerate flavours of sea quark  ($\nf=2$) 
where the same discretization scheme has been employed for the sea and light 
valence quarks.
In order to carry out the calculation with the computational resources 
available, the heavy quarks are treated using an effective field theory 
approach, and the light quark and  gluon fields actions are improved to 
minimize the discretization error.
The calculation is also made in the quenched ($\nf=0$) case 
since these decay constants have not been calculated before 
with this combination of actions. 

Our quenched continuum estimates for the decay constants are somewhat large
compared with the current summary of quenched results.
We note, however, that 
we have included an extra $1/M$ correction to the current, 
which was considered small and hence ignored in the previous calculations. 
We have demonstrated that this is not entirely true, giving a contribution 
of the order of $+3$\% to $\fb$ and $\fbs$ and $+6$\% to $\fd$ and $\fds$.
Allowing for the additional uncertainties from the use of one-loop 
perturbative renormalization factors, we consider our quenched results 
to be consistent with the previous results.

In comparing our $\nf=0$ and $\nf=2$ results we see that $\fb$ and $\fbs$  
for $\nf=2$ are significantly larger than the $\nf=0$ results by 2-3 
standard deviations, indicating a shift of 10--15\%.
On the other hand, the same cannot be said for the decay constants $\fd$ 
and $\fds$. 
It is encouraging that our prediction $f_{D_s}=267(^{+24}_{-21})$~MeV  with
$\nf=2$, where the total error is obtained by quadrature, 
is consistent with recent experiments.
In conjunction with the available experimental data, our values for 
the $\nf=2$  $B$ meson decay constants $f_{B_d}=208(15)$~MeV and 
$f_{B_s}=250(^{+18}_{-16})$~MeV 
are consistent with the hypothesis that the Wolfenstein parameter 
$\rho$ \cite{Wolfenstein} is positive. 
Given our results for $\nf=0$ and $\nf=2$, it is reasonable to think that 
additional flavours of sea quarks will increase $\fb$ and $\fbs$ still 
further,  which in turn favours a positive value for $\rho$ even more.

The unsatisfatory aspect of our results is a sizable 
variation of the decay constants with lattice spacing.
A possible origin of this problem is a necessity to include 
$O(a)$ and higher improvement terms in the axial vector current.  
Higher order corrections in the renormalization constants may also 
be important at the coarse lattice spacings of $a^{-1}\approx 1$--$2$~GeV 
explored in the present simulation.  The study of these issues
is clearly needed to consolidate the results for $\nf=2$ 
and further explore the final goal of predicting the 
heavy-light decay constants for the realistic spectrum of 
dynamical sea quarks.

\section*{Acknowledgments}
The authors would like to express their thanks to K.-I.~Ishikawa for
providing expressions for the axial renormalization coefficient.  
This work is supported in part by the Grants-in-Aid of Ministry of
Education (Nos. 09304029, 10640246, 10640248, 10740107, 11640250, 11640294,
11740162).  TM and AAK are supported by the
JSPS Research for the Future Program (Project No. JSPS-RFTF 97P01102).
SE, KN and HPS are JSPS Research Fellows.


\newpage

\newpage 

\newpage

\newpage

\newpage

\newpage

\newpage

\newpage

\begin{table}
\begin{center}
  \begin{tabular}{ccccccc}
    $\beta$ & $c_{SW}$ &  $K_{sea}$ & 
    $m_{PS}/m_V$ & 
    \#traj & 
    $K_s(K)^{-1}$ & $K_s(\phi)^{-1}$ \\ 
    \hline
    1.8 & 1.60 & 0.1409 & 0.807(1) & 5680(6250) & 6.929(3) & 7.037(6) \\   
        &      & 0.1430 & 0.753(1) & 5200(5200) & 6.945(4) & 7.045(8) \\   
        &      & 0.1445 & 0.694(2) & 6530(7000) & 6.956(3) & 7.044(7) \\   
        &      & 0.1464 & 0.547(4) & 4070(5250) & 6.969(4) & 7.028(8) \\   
    \hline                                
    1.95 & 1.53& 0.1375 & 0.804(1) & 6810(7000) & 7.144(2) & 7.190(3) \\   
         &     & 0.1390 & 0.752(1) & 5000(7000) & 7.154(2) & 7.196(3) \\   
         &     & 0.1400 & 0.690(1) & 6800(7000) & 7.164(2) & 7.202(3) \\
         &     & 0.1410 & 0.582(3) & 4870(7000) & 7.166(2) & 7.195(4) \\   
    \hline                                
    2.1 & 1.47 & 0.1357 & 0.810(2) & 1990(4000) & 7.283(3) & 7.306(6) \\   
        &      & 0.1367 & 0.757(3) & 2000(4000) & 7.282(2) & 7.298(4) \\   
        &      & 0.1374 & 0.693(3) & 1910(4000) & 7.285(2) & 7.299(4) \\   
        &      & 0.1382 & 0.571(6) & 1945(4000) & 7.285(3) & 7.299(5) \\   
  \end{tabular} 
  \caption{Simulation parameters for $\nf=2$ lattices used
    in the heavy quark calculation. For the number of trajectories those 
    in parentheses show the full ensemble generated.
    }
  \label{table:lattice-params-nf2}
\end{center}
\end{table}

\begin{table}
\begin{center}
  \begin{tabular}{ccc}
    $\beta$ &  $g^2_{\overline{MS}}(1/a)$ &
    $a$ (GeV$^{-1}$) \\
    \hline
    1.8 &  3.162 &1.090(11) \\
    1.95 &  2.812 & 0.7882(85) \\
    2.1 & 2.562 &  0.559(11) \\
\end{tabular}
  \caption{Chirally extapolated  parameters for $\nf=2$ lattices used
    in the heavy quark calculation. The lattice spacing is fixed by $\rho$ 
    meson mass.
    }
  \label{table:lattice-params-chiral-nf2}
\end{center}
\end{table}


\begin{table}
\begin{center}
  \begin{tabular}{ccccccccc}
    $\beta$ & $c_{SW}$ & 
    $g^2_{\overline{MS}}(1/a)$ &
    $a$ (GeV$^{-1}$) & \#conf & $\# K_h$ & $\# K_l$ & 
    $K_s(K)^{-1}$ & $K_s(\phi)^{-1}$ \\
    \hline
    2.187 & 1.439 & 2.809 & 1.017(10) & 200 & 7 & 2 & 7.274(4) & 7.326(8) \\   
    2.214 & 1.431 & 2.767 & 0.966(10) & 200 & 7 & 2 & 7.293(4) & 7.340(8) \\   
    2.247 & 1.422 & 2.716 & 0.917(9)  & 200 & 7 & 2 & 7.316(4) & 7.356(7) \\   
    2.281 & 1.412 & 2.664 & 0.896(10) & 220 & 7 & 2 & 7.348(4) & 7.395(8) \\   
    2.281 & 1.398 & 2.587 & 0.829(8)  & 200 & 6 & 3 & 7.379(3) & 7.420(6) \\   
    2.416 & 1.378 & 2.477 & 0.734(9)  & 190 & 8 & 2 & 7.415(4) & 7.452(7) \\  
    2.456 & 1.370 & 2.432 & 0.674(6)  & 190 & 8 & 2 & 7.422(2) & 7.449(4) \\  
    2.487 & 1.363 & 2.401 & 0.652(7)  & 200 & 8 & 2 & 7.434(3) & 7.462(5) \\  
    2.528 & 1.355 & 2.349 & 0.612(6)  & 195 & 8 & 2 & 7.446(2) & 7.471(4) \\  
    2.575 & 1.345 & 2.298 & 0.574(6)  & 200 & 8 & 3 & 7.458(2) & 7.480(4) \\  
  \end{tabular} 
  \caption{Simulation parameters for $\nf=0$.
    The lattice size employed is 
    $16^3 \times 32$ for $\beta$=2.187--2.281 and
    $24^3 \times 48$ for $\beta$=2.416--2.575.
    The lattice spacing is fixed by $\rho$ meson mass.} 
  \label{table:lattice-params-nf0}
\end{center}
\end{table}

\begin{table}
\begin{center}
  \begin{tabular}{c c  c  c  c   }
    $\beta$ & $\nf$ & $K_{sea}$ & $A$ & $B$ \\
    \hline
    1.8 & 2 & 0.1409 & 1.09 & 0.91 \\
    1.8 & 2 & 0.1430 & 1.09 & 0.91 \\
    1.8 & 2 & 0.1445 & 1.09 & 0.91 \\
    1.8 & 2 & 0.1430 & 1.09 & 0.91 \\
    \hline
    1.95 & 2 & 0.1375 & 1.28 & 0.58 \\
    1.95 & 2 & 0.139 & 1.28 & 0.58 \\
    1.95 & 2 & 0.140 & 1.23 & 0.71 \\
    1.95 & 2 & 0.141 & 1.27 & 0.6 \\
    \hline
    2.1 & 2 & 0.1357 & 1.28 & 0.54 \\
    2.1 & 2 & 0.1367 & 1.28 & 0.54 \\
    2.1 & 2 & 0.1374 & 1.28 & 0.54 \\
    2.1 & 2 & 0.1382 & 1.28 & 0.54 \\
    \hline
    \hline
    2.187 & 0 & - & 1.28 & 0.58 \\
    2.214 & 0 & - & 1.28 & 0.58 \\
    2.247 & 0 & - & 1.28 & 0.58 \\
    2.281 & 0 & - & 1.28 & 0.58 \\
    2.416 & 0 & - & 1.28 & 0.54 \\
    2.456 & 0 & - & 1.28 & 0.54 \\
    2.487 & 0 & - & 1.28 & 0.54 \\
    2.528 & 0 & - & 1.28 & 0.54 \\
    2.575 & 0 & - & 1.28 & 0.54 
  \end{tabular} 
  \caption{Smearing parameters used. 
    The gauge-fixed smearing function takes the form 
    $A\exp{(-Br)}$
    }
  \label{table:smearing-params}
\end{center}
\end{table}

\newpage

\begin{table}
\begin{center}
  \begin{tabular}{c c  c  c  c  c  c }
    $\beta$ &  $A_0$ &  $A_1$  
    & $B_0$ & $B_1$  & $C_0$  & $\chi^2/\mbox{d.o.f.}$  \\
    \hline
    2.187 & 0.670(27) & 1.10(6)  & -0.79(8)  & -1.17(9) & 0.485(60) & 1.9/9 \\
    2.214 & 0.597(29) & 1.01(9) & -0.66(8)  & -0.95(13) & 0.373(55) & 1.3/9 \\
    2.247 & 0.556(25) & 1.00(8) & -0.57(7)  & -0.88(11) & 0.296(51) & 0.7/9 \\
    2.281 & 0.480(25) & 1.00(13) & -0.43(6)  & -0.86(17) & 0.196(36) & 0.9/9 \\
    2.334 & 0.412(22) & 0.93(10) & -0.36(5) & -0.78(14) & 0.173(31) & 2.1/16\\
    2.416 & 0.319(12) & 0.84(5)  & -0.20(2)  & -0.60(5) & 0.055(10)  & 0.4/11\\
    2.456 & 0.317(2) & 0.52(24) & -0.21(3)  & -0.31(22) & 0.065(7) & 0.7/11 \\
    2.487 & 0.264(10) & 0.78(4) & -0.16(2)  & -0.47(4) & 0.044(6) & 0.3/11 \\
    2.528 & 0.237(16) & 0.82(34) & -0.13(1)  & -0.50(23) & 0.032(4) & 1.2/11\\
    2.575 & 0.235(9) & 0.67(5) & -0.14(1)  & -0.36(4) & 0.043(4) & 3.8/11
  \end{tabular} 
  \caption{Chiral HQET fit parameters for $\nf=0$.}
  \label{table:chiral-hqet-params-nf0}
\end{center}
\end{table}

\begin{table}
\begin{center}
  \begin{tabular}{c c  c  c   c  }
    $\beta$ &  $A_0$ &  $B_0$ & $C_0$ & $\chi^2/\mbox{d.o.f.}$  \\
    \hline
    2.187 & 0.785(23) & -0.89(7) & 0.456(63) & 0.7/4 \\
    2.214 & 0.699(22) & -0.75(7) & 0.366(57) & 0.6/4\\
    2.247 & 0.650(20) & -0.65(6) & 0.293(51) & 0.3/4\\
    2.281 & 0.583(15) & -0.54(5) & 0.228(38) & 0.8/4\\
    2.334 & 0.492(13) & -0.42(4) & 0.172(32) & 0.4/4\\
    2.416 & 0.365(15) & -0.24(3) & 0.063(22) & 0.006/4 \\
    2.456 & 0.357(10) & -0.25(2) & 0.076(11) & 0.1/4 \\
    2.487 & 0.315(8) & -0.19(2) & 0.045(8) & 0.03/4 \\
    2.528 & 0.288(12) & -0.16(2) & 0.030(13) & 0.2/3\\
    2.575 & 0.273(7) & -0.17(1) & 0.051(6) & 0.4/3 
  \end{tabular} 
  \caption{Strange (defined from the $K$) HQET fit parameters for $\nf=0$.}
  \label{table:strange-K-hqet-params-nf0}
\end{center}
\end{table}

\begin{table}
\begin{center}
  \begin{tabular}{c c  c  c  c   }
    $\beta$ &  $A_0$ &  $B_0$ & $C_0$ & $\chi^2/\mbox{d.o.f.}$  \\
    \hline
    2.187 & 0.814(22) & -0.95(7) & 0.492(60) & 0.7/4 \\
    2.214 & 0.723(20) & -0.80(6) & 0.392(55)& 0.7/4 \\
    2.247 & 0.667(20) & -0.67(6) & 0.300(52) & 0.3/4 \\
    2.281 & 0.611(15) & -0.59(5) & 0.257(46) & 1.1/4\\
    2.334 & 0.512(13) & -0.45(4) & 0.188(28) & 0.5/4 \\
    2.416 & 0.380(13) & -0.25(3) & 0.063(18) & 0.006/4\\
    2.456 & 0.362(7) & -0.25(2) & 0.076(10) & 0.1/4\\
    2.487 & 0.326(8) & -0.20(1) & 0.047(8) & 0.02/4 \\
    2.528 & 0.301(20) & -0.17(4) & 0.036(22) &  0.07/4 \\
    2.575 & 0.282(6) & -0.18(1) & 0.052(6) & 0.3/4 
  \end{tabular} 
  \caption{Strange (defined from the $\phi$) HQET fit parameters for $\nf=0$.}
  \label{table:strange-phi-hqet-params-nf0}
\end{center}
\end{table}

\begin{table}
\begin{center}
  \begin{tabular}{c c  c  c  c  c  c  c  }
    $\beta$ &  $A_0$ &  $A_1$ & $A_2$  
    & $B_0$ & $B_1$  & $C_0$ & $\chi^2/\mbox{d.o.f.}$ \\
    \hline
    1.80 & 0.95(2) & 0.48(4) & -0.10(2) & -1.27(6) & -0.31(3) & 0.76(3) & 30.4/26 \\
    1.95 & 0.51(2) & 0.61(6) & -0.14(4) & -0.72(8) & -0.49(7) & 0.58(8) & 18.7/26 \\
    2.10 & 0.25(1) & 0.56(9) & -0.34(15) & -0.18(1) & -0.25(2) & 0.06(5)  & 7.8/26 
  \end{tabular} 
  \caption{Chiral HQET fit parameters for $\nf=2$.}
  \label{table:chiral-hqet-params-nf2}
\end{center}
\end{table}

\begin{table}
\begin{center}
  \begin{tabular}{c c  c  c  c  c  c c  }
    $\beta$ &  $A_0$ &  $A_1$ & $A_2$  
    & $B_0$ & $B_1$  & $C_0$ & $\chi^2/\mbox{d.o.f.}$ \\
    \hline
    1.80 & 1.09(2) & 2.57(28) & -6.83(119) & -1.43(4) & -1.14(18) & 0.78(3) & 33.4/26 \\
    1.95 & 0.60(2) & 2.58(37) & -4.71(225) & -0.79(7) & -2.06(43) & 0.57(6) & 19.8/26 \\
    2.10 & 0.30(1) & 1.32(36) & -3.86(287) & -0.20(1) & -0.58(10) & 0.06(5) & 6.2/26
  \end{tabular} 
  \caption{Strange (defined from the $K$) HQET fit parameters for $\nf=2$.}
  \label{table:strange-K-hqet-params-nf2}
\end{center}
\end{table}

\begin{table}
\begin{center}
  \begin{tabular}{c c  c  c  c  c  c  c }
    $\beta$ &  $A_0$ &  $A_1$ & $A_2$  
    & $B_0$ & $B_1$  & $C_0$ & $\chi^2/\mbox{d.o.f.}$ \\
    \hline
    1.80 & 1.13(2) & 2.64(26) & -7.25(156) & -1.49(3) & -1.13(18) & 0.81(3) & 30.9/26 \\
    1.95 & 0.62(3) & 2.65(37) & -5.55(217) & -0.82(8) & -1.97(42) & 0.59(7) & 11.2/26\\
    2.10 & 0.31(1) & 1.22(36) & -2.69(282) & -0.21(1) & -0.63(9) & 0.07(4) & 4.5/26 
  \end{tabular} 
  \caption{Strange (defined from the $\phi$) HQET fit parameters for $\nf=2$.}
  \label{table:strange-phi-hqet-params-nf2}
\end{center}
\end{table}

\begin{table}
\begin{center}
  \begin{tabular}{ccccccc}
    $n_f$ & $\beta$ & 
    $\fb$        (GeV) & 
    $\fbs(K)$    (GeV) & 
    $\fbs(\phi)$ (GeV) & 
    $\fbs(K)/\fb$ & 
    $\fbs(\phi)/\fb$ \\
    \hline
    2 & 1.80  & 0.287(7) & 0.331(5) & 0.340(5) & 1.152(19) & 1.181(26)\\   
    2 & 1.95  & 0.234(8) & 0.276(7) & 0.283(8)& 1.179(43) & 1.211(45)\\   
    2 & 2.10  & 0.208(10)& 0.250(10)& 0.258(10)& 1.203(29) & 1.241(36)\\   
    \hline
    0 & 2.187 & 0.229(7) & 0.268(5) & 0.276(5) & 1.171(13) & 1.121(16)\\   
    0 & 2.214 & 0.220(8) & 0.258(5) & 0.265(5) & 1.169(20) & 1.202(25)\\   
    0 & 2.247 & 0.223(6) & 0.260(5) & 0.266(5) & 1.165(16) & 1.194(19)\\   
    0 & 2.281 & 0.204(7) & 0.244(4) & 0.254(3) & 1.196(31) & 1.243(40)\\   
    0 & 2.334 & 0.195(7) & 0.232(4) & 0.240(4) & 1.186(29) & 1.227(35)\\   
    0 & 2.416 & 0.188(6) & 0.212(5) & 0.220(5) & 1.126(11) & 1.169(15)\\   
    0 & 2.456 & 0.204(11)& 0.228(4) & 0.232(3) & 1.111(58) & 1.103(39)\\   
    0 & 2.487 & 0.184(5) & 0.217(4) & 0.226(3) & 1.182(14) & 1.221(18)\\   
    0 & 2.528 & 0.182(11) & 0.219(6) & 0.228(9)& 1.190(104)& 1.235(130)\\   
    0 & 2.575 & 0.192(6) & 0.221(4) & 0.227(4) & 1.151(18) & 1.183(21)\\   
  \end{tabular} 
  \caption{
    Decay constants $\fb$ and $\fbs$ at each bare gauge
    coupling.
    } 
  \label{table:final-results-B}
\end{center}
\end{table}

\begin{table}
\begin{center}
  \begin{tabular}{ccccccc}
    $n_f$ & $\beta$ &
    $\fd$        (GeV) &
    $\fds(K)$    (GeV) &
    $\fds(\phi)$ (GeV) &
    $\fds(K)/\fd$ & 
    $\fds(\phi)/\fd$  \\   
    \hline
    2 & 1.80  & 0.301(9) & 0.346(7) & 0.352(6) & 1.151(18)& 1.168(24)\\   
    2 & 1.95  & 0.284(13)& 0.312(11)& 0.316(11)& 1.097(40)& 1.111(39)\\
    2 & 2.10  & 0.225(14)& 0.267(13)& 0.277(13)& 1.182(39)& 1.223(47)\\   
    \hline
    0 & 2.187 & 0.258(4) & 0.295(3) & 0.300(3) & 1.143(7) & 1.165(8) \\   
    0 & 2.214 & 0.247(4) & 0.284(3) & 0.290(3) & 1.149(9) & 1.172(12)\\   
    0 & 2.247 & 0.250(4) & 0.287(3) & 0.293(3) & 1.146(8) & 1.169(10)\\   
    0 & 2.281 & 0.236(4) & 0.274(2) & 0.282(2) & 1.160(17)& 1.191(22)\\   
    0 & 2.334 & 0.229(4) & 0.263(3) & 0.269(3) & 1.148(12)& 1.175(14)\\   
    0 & 2.416 & 0.222(4) & 0.246(3) & 0.253(3) & 1.104(9) & 1.135(11)\\   
    0 & 2.456 & 0.229(8) & 0.254(3) & 0.257(3) & 1.129(73)& 1.119(46)\\   
    0 & 2.487 & 0.215(3) & 0.249(2) & 0.256(2) & 1.159(7) & 1.188(9) \\   
    0 & 2.528 & 0.215(10)& 0.250(4) & 0.256(6) & 1.133(67)& 1.159(81)\\   
    0 & 2.575 & 0.215(4) & 0.245(3) & 0.251(3) & 1.138(13)& 1.163(16)\\   
  \end{tabular} 
  \caption{
    Decay constants $\fd$ and $\fds$ at each bare gauge
    coupling.
    } 
  \label{table:final-results-D}
\end{center}
\end{table}

\begin{table}
\begin{center}
  \begin{tabular}{cccc}
    $\beta$ & 1.8 & 1.95 & 2.1 \\
    \hline
    $(\Lambda_{QCD}/{M_B})^2$, $(\Lambda_{QCD}/{M_D})^2$ & 
    $<$1\%, 3\% & $<$1\%, 3\% & $<$1\%, 3\% \\
    $\alpha_s^2$              &  5\% & 3\% & 3\% \\
    $(a\Lambda_{QCD})^2$      & 11\% & 6\% & 3\% \\  
    $\alpha_s a\Lambda_{QCD}$ &  6\% & 4\% & 2\% \\  
    $\alpha_s\Lambda_{QCD}/M_B$,  
    $\alpha_s\Lambda_{QCD}/M_D$ &
    1\%, 3\% & 1\%, 3\% & 1\%,3\% \\  
    \hline
    total (linear) & 23\%, 28\% & 14\%, 19\% & 9\%, 14\% \\
    total (quadratic) & 14\%, 14\% & 8\%, 9\% & 5\%, 6\% \\  
  \end{tabular} 
  \caption{An estimate of systematic errors for $\nf=2$. 
    $\Lambda_{QCD}$ is taken to be 300 MeV.
By convention, we use the quadratically summed error.
    } 
  \label{table:lattice-sys-nf2}
\end{center}
\end{table}

\begin{table}
\begin{center}
  \begin{tabular}{cccc}
    $\beta$ & 2.187 & 2.416 & 2.575 \\
    \hline
    $(\Lambda_{QCD}/M_B)^2$, $(\Lambda_{QCD}/M_D)^2$ 
    & $<$1\%, 3\% & $<$1\%, 3\% & $<$1\%, 3\% \\
    $\alpha_s^2$              & 3\% & 3\% & 4\% \\  
    $(a\Lambda_{QCD})^2$      & 9\% & 5\% & 3\% \\  
    $\alpha_s a\Lambda_{QCD}$ & 5\% & 3\% & 3\% \\  
    $\alpha_s\Lambda_{QCD}/M_B$, $\alpha_s\Lambda_{QCD}/M_D$ & 
    1\%, 3\% & 1\%, 3\% &  1\%,3\% \\  
    \hline
    total (linear) & 18\%, 23\% & 12\%, 17\% & 11\%, 16\% \\
    total (quadratic) & 11\%, 12\% & 7\%, 8\% & 5\%, 7\% \\  
  \end{tabular} 
  \caption{An estimate of systematic errors for $\nf=0$. 
    $\Lambda_{QCD}$ is taken to be 300 MeV. By convention, we use
the quadratically summed error.
    } 
  \label{table:lattice-sys-nf0}
\end{center}
\end{table}


\newpage

\begin{thebibliography}{99}

\bibitem{Buras:1990fn}
  A.J.~Buras, M.~Jamin and P.H.~Weisz,
  Nucl. Phys. \textbf{B347} (1990) 491.

\bibitem{ALEPH}
  ALEPH collaboration, ALEPH 2000-062, contributed paper to 
  XXXth International Conference on High Energy Physics,
  July 27 - August 2, 2000, Osaka, Japan.

\bibitem{CLEO}
  CLEO collaboration (M.~Chadha \textit{et al.}),
  Phys. Rev. \textbf{D58}, 32002 (1998).

\bibitem{Hashimoto:1999bk}
  S.~Hashimoto,
  Nucl. Phys. B (Proc. Suppl.) \textbf{83-84} (2000) 3.

\bibitem{Draper_lat98}
  T.~Draper, Nucl. Phys. B (Proc. Suppl.) \textbf{73} (1999)
  43.

\bibitem{Aoki:2000yr}
  CP-PACS collaboration (S. Aoki \textit{et al.}),
  Phys. Rev. Lett. \textbf{84} (2000) 238.

\bibitem{Booth:1995hx}
  M.J.~Booth,
  Phys. Rev. \textbf{D51} (1995) 2338.

\bibitem{Sharpe:1996qp}
  S.R.~Sharpe and Y.~Zhang,
  Phys. Rev. \textbf{D53} (1996) 5125.

\bibitem{Bernard:1998xi}
  C.~Bernard \textit{et al.},
  Phys. Rev. Lett. \textbf{81} (1998) 4812;
  Nucl. Phys. B (Proc. Suppl.) \textbf{83-84} (2000) 289.

\bibitem{Collins:1999ff}
  S.~Collins, C.T.~Davies, U.M.~Heller, A.~Ali Khan,
  J.~Shigemitsu, J.~Sloan and C.~Morningstar, 
  Phys. Rev. \textbf{D60} (1999) 074504.

\bibitem{ref:clover}
  B.~Sheikholeslami and R.~Wohlert, 
  Nucl. Phys. \textbf{B259} (1985) 572.

\bibitem{El-Khadra:1997mp}
  A.X.~El-Khadra, A.S.~Kronfeld and P.B.~Mackenzie,
  Phys. Rev. \textbf{D55} (1997) 3933.

\bibitem{ref:Iwasaki83}
  Y.~Iwasaki, 
  Nucl. Phys. \textbf{B258} (1985) 141;
  Univ.\ of Tsukuba report UTHEP-118 (1983), unpublished.

\bibitem{csw-one-loop}
S.~Aoki {\it et al.} Nucl. Phys. {\bf B540}, 501 (1999).

\bibitem{Kaneko}
  CP-PACS collaboration (S. Aoki \textit{et al.}),
  Phys. Rev. \textbf{D60} (1999) 114508.

\bibitem{HQET}
  E.~Eichten and B.~Hill, Phys. Lett. \textbf{B234} (1990) 511;
  H.~Georgi, Phys. Lett. \textbf{B240} (1990) 447.

\bibitem{static}
  E.~Eichten, 
  Nucl. Phys. B (Proc. Suppl.) \textbf{4} (1988) 170.

\bibitem{NRQCD-1}
  B.A.~Thacker and G.P.~Lepage, 
  Phys. Rev. \textbf{D43} (1991) 196.

\bibitem{NRQCD-2}
  G.P.~Lepage \textit{et al.}, 
  Phys. Rev. \textbf{D46} (1992) 4052. 

\bibitem{Mertens_et_al}
  B.P.G.~Mertens, A.S.~Kronfeld and A.X.~El-Khadra,
  Phys. Rev. \textbf{D58} (1998) 034505.

\bibitem{Kuramashi}
  Y. Kuramashi, 
  Phys. Rev. \textbf{D58} (1998) 034507.

\bibitem{AHIO}
  S. Aoki, S. Hashimoto, K.-I. Ishikawa and T. Onogi, unpublished (1997).

\bibitem{LM}
  G.P.~Lepage and P.B.~Mackenzie,
  Phys. Rev. \textbf{D48} (1993) 2250.

\bibitem{BLS} 
  C. W. Bernard, J. N. Labrenz and A. Soni, 
  Phys. Rev. \textbf{D49} (1994) 2536.

\bibitem{Aoki:1998ji}
  JLQCD collaboration (S.~Aoki \textit{et al.}),
  Phys. Rev. Lett. \textbf{80} (1998) 5711.

\bibitem{El-Khadra:1998hq}
  A.X.~El-Khadra, A.S.~Kronfeld, P.B.~Mackenzie, S.M.~Ryan
  and J.N.~Simone, 
  Phys. Rev. \textbf{D58} (1998) 014506.

\bibitem{Burkhalter:1998wu}
  R.~Burkhalter,
  Nucl. Phys. B (Proc. Suppl.) \textbf{73} (1999) 3.



\bibitem{CP-PACS_lightquark}
  CP-PACS Collaboration (A. Ali Khan {\it et al.}), 
  hep-lat/0004010 (2000).

\bibitem{CP-PACS_lighthadron}
  CP-PACS Collaboration (A. Ali Khan {\it et al.}), 
  in preparation.

\bibitem{Shanahan:1997pk}
  H.P.~Shanahan {\it et al.},
  [UKQCD Collaboration],
  Phys. Rev. \textbf{D55} (1997) 1548.

\bibitem{Aoki:1998ar}
  S.~Aoki, K.~Nagai, Y.~Taniguchi and A.~Ukawa,
  Phys. Rev. \textbf{D58}, 074505 (1998).

\bibitem{Ishikawa}
  K.-I. Ishikawa, T. Onogi and N. Yamada, 
  Nucl. Phys.(Proc.Suppl.) {\bf 83} (2000) 301; 
  K.-I.~Ishikawa, private communication.

\bibitem{Wolfenstein}
  L. Wolfenstein, Phys. Rev. Lett., {\bf 51}, 1945, (1983).

 

\end{thebibliography}
\end{document}